\documentclass[aps,prl,portrait,twocolumn,superscriptaddress,longbibliography,noeprint]{revtex4-2}

\usepackage{graphicx}
\usepackage[normalem]{ulem}
\usepackage{color}

\usepackage{xspace}
\usepackage{amsmath}
\usepackage{siunitx}
\usepackage{miller}

\usepackage[version=3]{mhchem}
\usepackage{natbib}
\usepackage{orcidlink}

\usepackage{pdfpages}
\makeatletter
\AtBeginDocument{\let\LS@rot\@undefined}
\makeatother

\newcommand{\nio}{Na$_2$IrO$_{3}$}
\newcommand{\lio}{Li$_2$IrO$_{3}$}
\newcommand{\arc}{$\alpha$-RuCl$_{3}$}

\begin{document}


\title{Direct evidence for anisotropic magnetic interaction in $\alpha$-RuCl$_3$ from polarized inelastic neutron scattering }

\author{Markus Braden\,\orcidlink{0000-0002-9284-6585}}
\email{braden@ph2.uni-koeln.de}
\affiliation{
	II. Physikalisches Institut, Universität zu Köln, Zülpicher Str. 77, D-50937 Köln, Germany\\
}

\author{Xiao Wang}%
\affiliation{J\"ulich Centre for Neutron Science JCNS at MLZ,
	Forschungszentrum J\"ulich,
	Lichtenbergstr. 1,
	D-85747 Garching, Germany\\
}
\affiliation{Institute of High Energy Physics, Chinese Academy of Sciences (CAS),
	Beijing 100049, China\\}
\affiliation{Spallation Neutron Source Science Center, Dongguan 523803, China\\}

\author{Alexandre Bertin\,\orcidlink{0000-0001-5789-3178}}
\affiliation{
	II. Physikalisches Institut, Universität zu Köln, Zülpicher Str. 77, D-50937 Köln, Germany\\
}

\author{Paul Steffens,\orcidlink{0000-0002-7034-4031}}
\affiliation{Institut Laue-Langevin, 71 avenue des Martyrs, CS 20156, 38042 Grenoble Cedex 9, France\\}%

\author{Yixi Su}
\email{y.su@fz-juelich.de}
\affiliation{J\"ulich Centre for Neutron Science JCNS at MLZ,
	Forschungszentrum J\"ulich,
	Lichtenbergstr. 1,
	D-85747 Garching, Germany\\
}%

\date{\today}

\begin{abstract}

Polarized neutron scattering experiments reveal the anisotropy of magnetic correlations in the candidate Kitaev material \arc . The anisotropy of the inelastic response at the magnetic Bragg positions is just opposite to 
the expectation for a simple Heisenberg model.  
Near the antiferromagnetic $q$ vector, there are no low-energy transversal magnon modes directly documenting the fully anisotropic and bond-directional character of the magnetic interaction in \arc .
However, other findings disagree
with a simple or strongly dominant Kitaev component.

\end{abstract}

\maketitle



The exactly solvable Kitaev model on a honeycomb lattice \cite{kitaev2006} has triggered enormous activities inspired 
by both the fundamental aspects and the possible impact for quantum-computing technologies \cite{trebst2022,takagi2019,winter2017b,broholm2020}. 
Although the honeycomb iridates \nio ~ and \lio ~ \cite{singh2010,chaloupka2010,jackeli2009} as well as the ruthenate \arc ~ \cite{plumb2014} were rapidly identified as candidate materials, none of these compounds could be safely established as a good realization of this unique model.

In the Kitaev model \cite{kitaev2006}, the magnetic interaction between two magnetic sites is bond directional, i.e. for a chosen bond in the hexagon of the honeycomb lattice only a single spin direction is coupled. The system contains three different types of nearest-neighbor bonds, labeled $x$, $y$, and $z$ bonds, that are rotated by 120 degrees against each other. In each bond type only one orthogonal spin component is coupled, the $x$, $y$ and $z$ component, respectively. In the honeycomb iridates and in \arc \ the lattice consists of edge-sharing octahedra \cite{trebst2022,takagi2019,winter2017b}. In such an edge-sharing octahedra pair the Kitaev interaction couples the components perpendicular to the plane formed by the two metal positions and the shared edge. Since the interaction in neighboring pairs connects orthogonal components, the Kitaev model cannot generate long-range order. Instead the ground state is a quantum spin liquid, for which precise predictions were made due to the exact solvability of this model. Most prominent is the prediction of the quantized thermal Hall effect \cite{kitaev2006}, but whether experiments confirm or refute this quantization remains matter of strong controversy \cite{kasahara2018,bruin2022,czajka2021}.
The possible realization of the long-searched quantum spin-liquid state in \arc \ 
and the nature of such a phase remain most challenging open questions \cite{broholm2020}.

The honeycomb iridates and \arc \ exhibit long-range antiferromagnetic order \cite{trebst2022,takagi2019,winter2017b,johnson2015,cao2016}, which implies the presence of additional magnetic coupling terms. The antiferromagnetic order in these most promising candidates is very similar and consists of antiferromagnetically stacked zigzag chains with parallel moments. Combining this antiferromagnetic structure type with DFT studies of the quantum chemistry \cite{mazin2012,foyevtsova2013}, a minimal model for the nearest-neighbor interaction was established containing the Kitaev interaction $J_\text{K}$, a Heisenberg term $J$ and two parameters for the symmetric anisotropic interaction, $\Gamma$ and $\Gamma'$ \cite{winter2017b,winter2017,liu2022}. The full nearest-neighbor interaction for the $x$ bond can be written as a quadratic form $S_i^T\cdot M \cdot S_j$ with the interaction matrix $M$:
$$ M= \left(\begin{matrix}
	J_\text{K}+J & \Gamma' & \Gamma' \\
	\Gamma' & J & \Gamma \\
	\Gamma' & \Gamma & J \\
\end{matrix}  \right). $$
For the $y$ and $z$ bonds the components need to be accordingly permutated. In addition, there are Heisenberg interaction terms between second- and third-nearest neighbors, $J_2$ and $J_3$, and eventually a single-ion anisotropy parameter. \arc ~ has been much better studied by neutron techniques \cite{banerjee2016,banerjee2017,banerjee2018,balz2019,ran2017,ran2022,do2017} than the Ir compounds, because Ru absorbs neutrons much less than Ir and due to the availability of large crystals. However, even for \arc \ an interaction model satisfactorily describing all experimental findings could not be established so far \cite{winter2017,liu2022,laurell2020,li2021,janssen2020,suzuki2021}. Furthermore, the  proposed collinear zigzag model for the antiferromagnetic structure possibly oversimplifies the real magnetic arrangement, see below. The studies of the magnetic correlations in \arc \ were hampered by pronounced sample dependencies related to the stacking of the RuCl$_3$ layers. Early studies observed a two-layer magnetic stacking of zigzag ordered planes \cite{janssen2020,johnson2015,ritter2016,cao2016}, an $ABAB$ scheme, which is consistent with a monoclinic nuclear structure in space group $C2/m$. In contrast more recent experiments on what now is considered being high-quality crystals, reveal a different stacking of the chemical layers described in space group $R\bar{3}$ \cite{park2024,kim2024,sarkis2024}. In this symmetry, neighboring layers shift along a bond direction by the length of one Ru-Ru distance, so that the third-nearest layer sits exactly on top of the initial one. Therefore, the magnetic structure cannot exhibit the two-layer $ABAB$ scheme but must follow a three-layer stacking  $ABCABC$ (or an even more complex one). The rhombohedral spacegroup is usually treated in a hexagonal setting that includes two centring translations and thus implies peculiar Bragg selection rules. Furthermore, rhombohedral crystals can be twinned, associated with the reverse and obverse setting of the centring translations. This twinning can be easily understood for \arc ; the two twins correspond to shifting the nearest layer in the $+z$ direction in either positive or negative bond direction. 

\begin{figure}[t]
	\centering
	\includegraphics[width=0.9992\columnwidth]{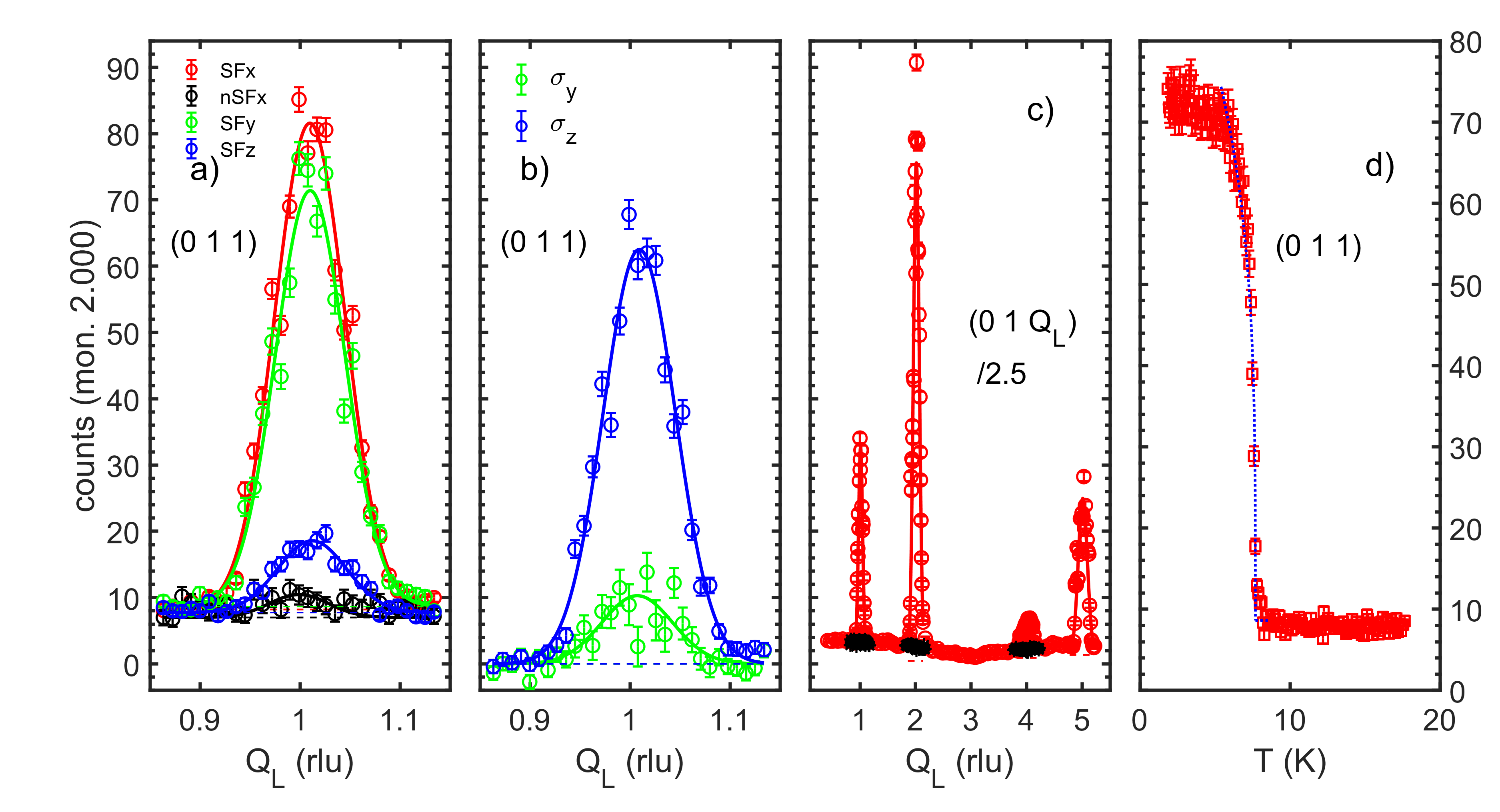}
	\caption{ { Elastic magnetic Bragg scattering studied with polarization analysis. Panel a) shows the three spin-flip channels SFx, SFy and SFz and the non-spin-flip channel NSFx for a scan across the (0,1,1) magnetic Bragg peak along $c^\star$ direction. From the spin-flip data we calculate the contributing moments
			in ${\bf y_n}$ and ${\bf z_n}$ directions (b). Panel (c) presents a long $Q_L$ scan across the magnetic peaks in the ordered (red) and paramagnetic (black) states. The temperature dependence of the (0,1,1) intensity is displayed in (d) with a power-law fit. }}
	\label{bragg}
\end{figure}

\begin{figure*}[t]
	\includegraphics[width=0.85\columnwidth]{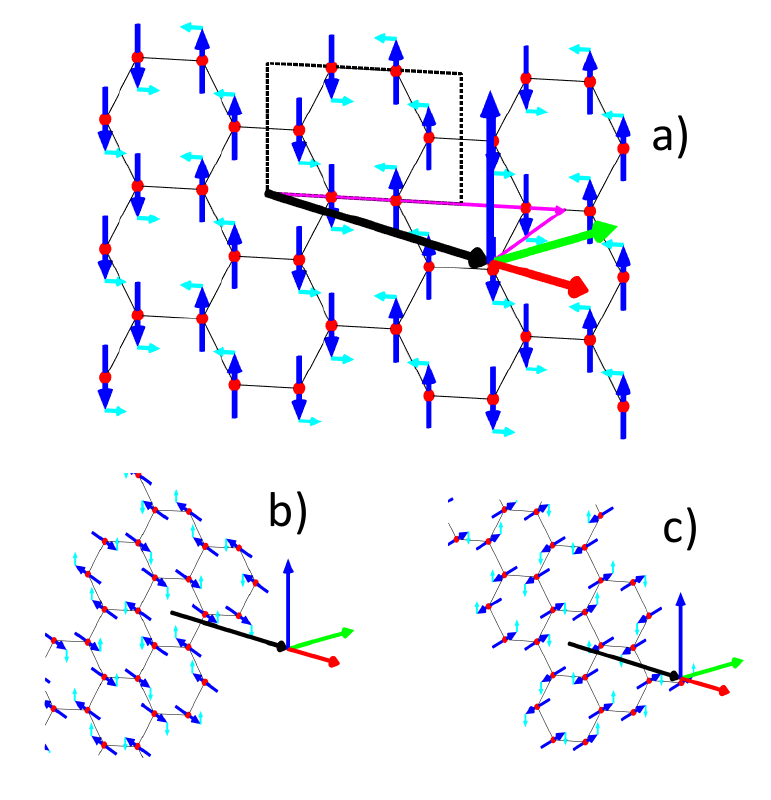}
	\includegraphics[width=0.99\columnwidth]{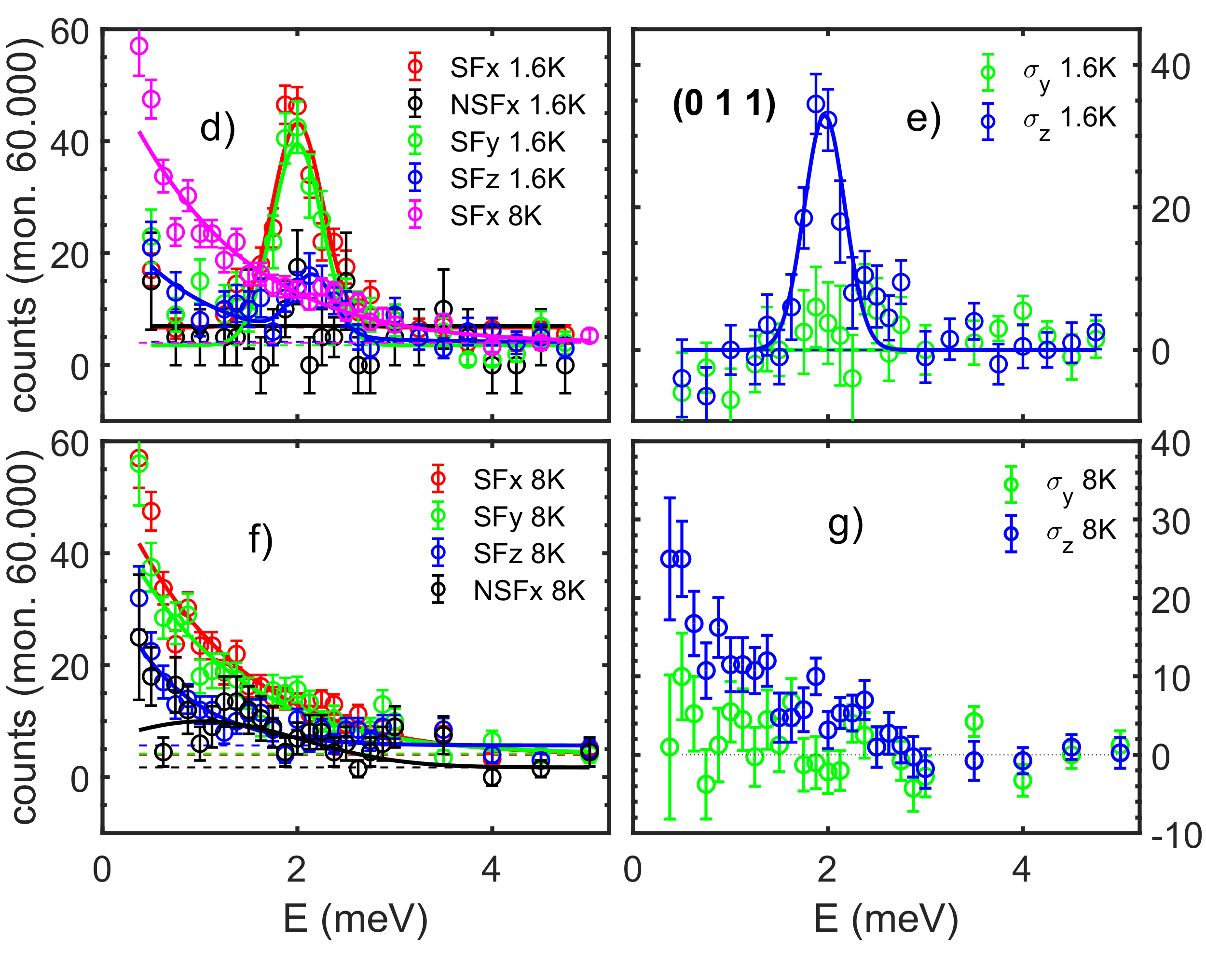}
	\caption{{Panel (a) illustrates the scattering geometry at the antiferromagnetic Bragg reflection (0,1,1): the horizontal scattering plane is spanned by [010] (in-plane) and the direction [001] perpendicular to the honeycomb layers. 
	Static ordered moments are drawn as dark blue arrows and the polarization of an usually expected low-energy transversal magnon is illustrated by light blue arrows.
	The axis vertical to the scattering plane (vertical in the drawing) corresponds to the perpendicular in-plane direction, ${\bf a}$, in real space.
	Panels (b) and (c) illustrate the scattering geometry for the two rotated magnetic domains, from which the
	zigzag order does not contribute to elastic intensity. 
	Panel (d) presents the polarized data of an energy scan at $Q$=(0,1,1) at 1.6\,K showing all SF channels and NSFx. 
	The SFx data taken at 8\,K are added to illustrate the rapid suppression of the sharp magnon mode. From these data the scattering contributions $\sigma_{y}$ and $\sigma_{z}$ were calculated (e). Panels (f,g) present the analogous data obtained above the N\'eel temperature at 8\,K. }}
	\label{fig:ins}
\end{figure*}

We use an orthorhombic setting of $R\bar{3}$ to keep the analogy with \nio \ and to avoid the complications of the non-orthogonal lattices. In the layer, $b\sim10$\AA \ parallel to a bond corresponds to an antiferromagnetic zigzag cell, and the axis $a\sim6$\AA \ is perpendicular to a bond and corresponds to a hexagonal in-plane parameter.
The $c$ parameter in this orthorhombic cell corresponds to three times the interlayer distance and is
thus identical to the parameter in the hexagonal setting of $R\bar{3}$.

For \nio \ evidence for bond-directional or at least anisotropic interaction was deduced from resonant elastic and inelastic X-ray scattering by analyzing the diffuse and high-energy response \cite{hwanchun2015,magnaterra2023}, but for \arc ~ there is only
indirect evidence by analyzing the magnetic excitations in anisotropic models \cite{chaloupka2015,winter2017,liu2022,laurell2020,li2021,janssen2020,maksimov2020,suzuki2021}. Polarized neutron scattering 
is able to directly sense the direction of a magnetic signal as only components perpendicular to the direction of neutron polarization analysis can contribute to spin-flip scattering, while the parallel components result in non-spin-flip scattering \cite{chatterji2005}. 
These rules add to the general selection condition that neutron scattering only senses the moments perpendicular to the scattering vector. Since one may arbitrarily choose the direction of the neutron polarization analysis and vary the scattering vector, one may fully characterize the anisotropy of the elastic and inelastic response \cite{qureshi2012,kunkemoeller2015,chatterji2005}. 
From such measurements on \arc \ we can deduce the fully anisotropic character
of the excitations, which directly points to the anisotropic microscopic interaction.

A large plate-shaped crystal of \arc \ ($\sim$0.7\,g mass) was grown by the chemical vapor transport method \cite{mi2021}. The crystal exhibits antiferromagnetic Bragg scattering corresponding to the triple-layer stacking with a high N\'eel temperature of 7.70\,K underlining its high quality. 
Polarized neutron scattering experiments were performed on the $Thales$ cold triple-axis spectrometer at the ILL \cite{a-rcl_thales}.  
We used focusing Heusler crystals as polarizing monochromator and analyzer, and a Helmholtz-coil setup enabled adjusting the neutron polarization direction at the sample. 
Flipping ratios determined on nuclear Bragg peaks (0,0,1) and (0,0,2) amounted to 27 and 20, respectively. 
The slightly lower value at the higher scattering angle arises from magnetic-guide-field crosstalks in the incoming and outgoing beams. 
To suppress higher order contamination, we inserted a velocity selector and a Beryllium filter with most scans performed at fixed k$_\text{f}$=1.50\,\AA$^{-1}$. 
The crystal was mounted in the [010]/[001] scattering geometry using a non-conventional setting, with $a$=5.98, $b$=10.32, and $c$=16.92\,\AA .

Antiferromagnetic Bragg peaks associated with the stacking of zigzag ordered planes appear at (0,1,$Q_L$) \cite{kim2024,park2024}; note that the $Q_L$=1, 4 and 2, 5 signals stem from distinct domains of the $R\bar{3}$ nuclear stacking, respectively. The scattering geometry is illustrated in Fig.~\ref{fig:ins} (a).  
Due to the finite out-of-plane component the scattering vector (0,1,$Q_L$) is canted  from the honeycomb layers. 
For the domain contributing to the Bragg scattering the bonds parallel to the in-plane component exhibit antiparallel spin alignment, while the bonds rotated by $\pm$120 degrees compared to those show parallel spins. 
Moments are supposed to be perpendicular to the antiferromagnetic bonds, i.e. they lay in the ${\bf a,c}$ plane in our notation and it is claimed that the moments form an angle of $\sim$31 degrees with the layers \cite{sears2020,kim2024b}. For simplicity the moments in  Fig.~\ref{fig:ins} (a) are drawn parallel to ${\bf a}$. In order to exploit the selection rules of polarized neutron scattering that are tied to the scattering vector it is convenient to introduce the coordinate system with ${\bf x_n}$ parallel to the scattering vector ${\bf Q}$, ${\bf z_n}$ vertical to the scattering plane and 
${\bf y_n}={\bf z_n}\times{\bf x_n}$, that is indicated in red-green-blue color coding. The ${\bf z_n}$ coordinate thus corresponds always exactly to ${\bf a}$, while ${\bf y_n}$ is parallel to the vertical direction ${\bf c}$ for $Q_L=0$ and rotates towards ${\bf b}$ for finite $Q_L$.

Figure~\ref{bragg} presents results of elastic scans across the antiferromagnetic Bragg positions. We focus on the three spin-flip (SF) channels along the above-given polarization directions ${\bf x_n,y_n,z_n}$
and record also the non-spin-flip (nSF) channel for polarization along ${\bf x_n}\parallel {\bf Q}$, because for this polarization direction the entire magnetic scattering appears in the SF channel. Note that nSF intensities in the two other directions contain also magnetic scattering polarized parallel to the respective direction. From the SF intensities one can calculate the magnetic signals polarized parallel to  ${\bf y_n}$( ${\bf z_n}$) by $\sigma_y=\text{SFx}-\text{SFy}$ ($\sigma_z=\text{SFx}-\text{SFz}$), which allows one to determine
these signals free of background. The data in Fig.~\ref{bragg} (a) show that SFy is almost as strong as SFx while SFz is weak. In the nSFx channel there is only a tiny signal due to the purely magnetic character of this Bragg peak and the high flipping ratio of the instrumental setup, see above. The calculated anisotropic components of the scattering intensities in Fig.~\ref{bragg} (b) thus reveal that the magnetic signal is mostly polarized along  ${\bf z_n}$ which corresponds to orthorhombic ${\bf a}$ as it is qualitatively expected. However, the suppression of the $\sigma_y$ component is much stronger than what is expected for a collinear model with a canting angle of $\sim$30 degrees \cite{sears2020,kim2024b}. 
Figure~\ref{bragg} (c) shows a long $Q_L$ scan at low temperature (SFx channel) and a few points taken in the paramagnetic phase confirming the full suppression of the magnetic Bragg signal. 
The temperature dependence of the magnetic intensity above T=5.4\,K was
fitted by a power law $\propto (T_\text{N}-T)^{2\beta}$ yielding $T_\text{N}$=7.7(1)\,K and $\beta$=0.14(1) consistent with the exact value for the 2D Ising model of $\beta$=0.125 \cite{leguillou1987,novotny1992} as well as with a previous analysis \cite{park2024}.

Figure~\ref{fig:ins} presents the polarization analysis of the inelastic response at the antiferromagnetic Bragg peak (0,1,1) in the ordered and in the paramagnetic states.
Note that a previous polarized experiment only revealed the magnetic character of the
peak in the ordered state but did not collect all SF channels \cite{ran2022}.
Panels (a-c) illustrate the geometry at this scattering vector for the three magnetic domains with the stacking vector
of the zigzag order being either parallel to (0,1,0) or rotated by $\pm$120\,degrees. The main message of our study becomes immediately evident. The inelastic low-energy response with a rather sharp peak at 2\,meV appears essentially in the same channel as the elastic Bragg scattering, which is opposite to
the expectation for a Heisenberg system with some weak anisotropy. For such a system, the low-energy excitations correspond to precession of the moments around the static component so that the inelastic 
response exhibits a transversal polarization and must thus appear in other channels compared to the
elastic Bragg scattering. The sharp mode thus cannot be attributed to the low-energy transversal magnon that attains finite energy from some moderate anisotropy as it is observed for example for the in-plane magnon in Ca$_2$RuO$_4$ \cite{kunkemoeller2015}. In \arc \ the anisotropy of magnetic interaction yields the dominant energy scale and thus completely changes the appearance of the magnon response. 
Transversal magnons are absent at low energy and must be pushed above the studied energy range, i.e. above $\sim$4\,meV. This is direct evidence for the dominance of magnetic anisotropy in this Kitaev candidate material.

\begin{figure}[t]
	\includegraphics[width=0.9995\columnwidth]{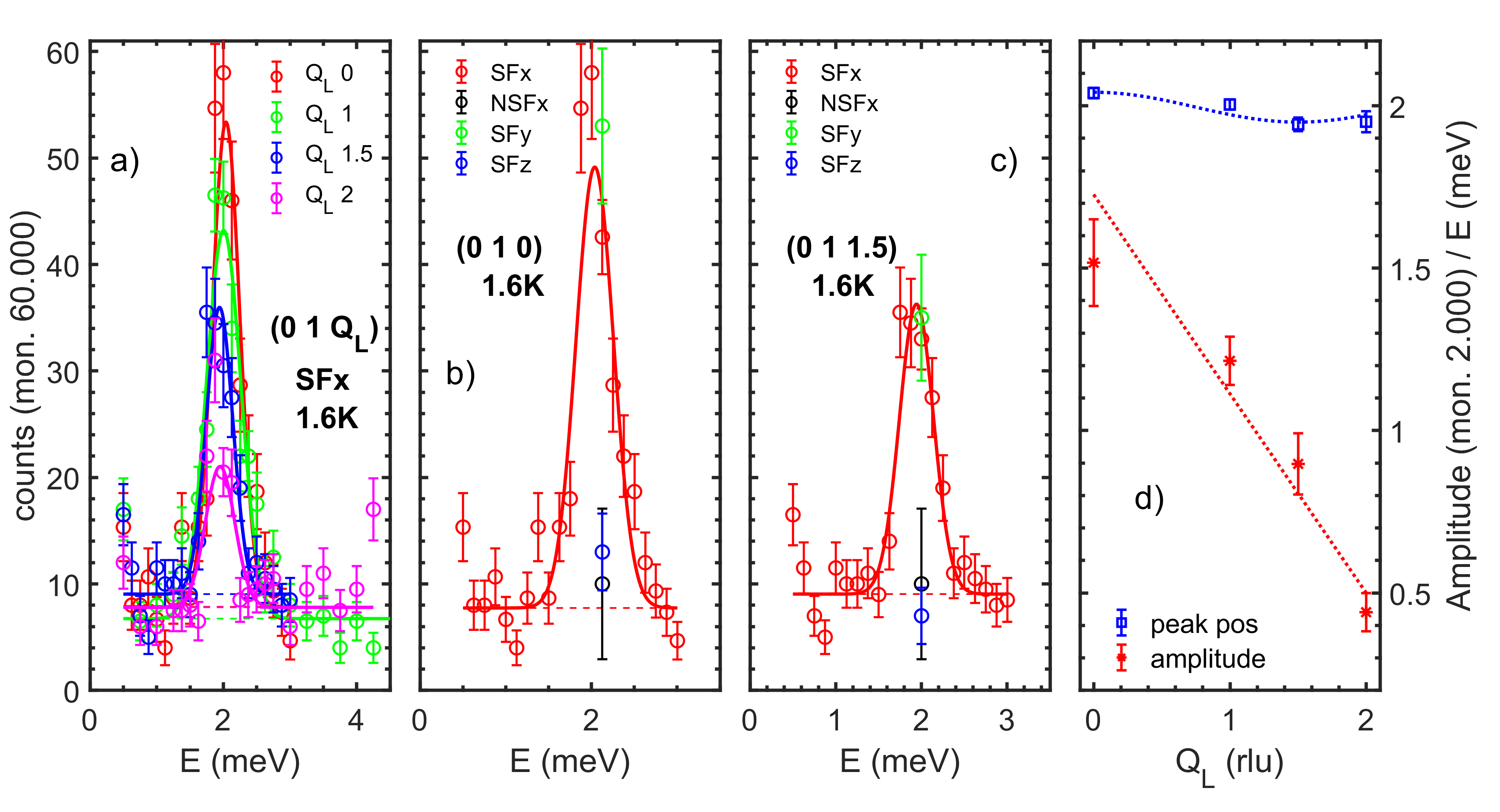}
	\caption{ { Results of polarized scans at the 2D antiferromagnetic zone centers (0,1,$Q_L$). 
			The SFx intensities are shown for different $Q_L$ values obtained at 1.6\,K. 
			The $Q_L$ dependencies of the amplitude and of the peak position are shown in panel (b). 
			Panels (c) and (d) show the data at (0,1,0) and (0,1,0.5), respectively, with a full polarization analysis near the peak.}}
	\label{fig:otherafm}
\end{figure}

The sharp peak at the antiferromagnetic scattering vector in \arc \  that has also been observed in previous unpolarized studies \cite{ran2017,balz2019} thus cannot be attributed to a transversal zone-center magnon. 
For special parameter ratios the extended Kitaev-Heisenberg model can be transformed to an isotropic Heisenberg model with Goldstone modes, and these modes are pushed to the zone-center in realistic non-constrained models \cite{chaloupka2015}.
We attribute the sharp feature at the magnetic Bragg peak to the response of domains, whose zigzag order is rotated against the
scattering vector and which thus do not contribute to the Bragg scattering at this $Q$=(0,1,1). In the single-$k$ picture of the zigzag order one such stacking has condensed into static order [Fig. 2a)] but the system should remain close to instability with respect to the condensation of these two rotated zigzag schemes  [Fig. 2b)  and c)]. The sharp low-energy peak in Fig. 2 thus corresponds to the soft correlations of the zigzag schemes
that are rotated by $\pm$120 degrees with respect to the static one.
The zigzag schemes are tight to the bonds, which are selected by the scattering vector $Q$. Therefore, the finding that the polarization of these excitations is nearly uniaxial and always perpendicular to the in-plane ${\bf Q}$ component directly confirms the bond-directional character of the
anisotropic magnetic interaction in \arc .

The interpretation of the magnetic response in the fully anisotropic picture is confirmed by linear spin-wave calculations. Implementing the parameter sets proposed by Winter et al., Maksimov et al. and by Liu et al. with the $SpinW$ program \cite{toth2015}, our calculations yield the dispersion in perfect agreement with these references \cite{winter2017,maksimov2020,liu2022}, see supplemental material \cite{suppl-mat}. At the antiferromagnetic Bragg vector there is no low-energy response in the mono-domain calculation. Such low-energy response only appears in the rotated directions from where it is folded to the Bragg position when simulating the response of a multidomain crystal.
These models, however, underestimate the energy of the initial transversal antiferromagnetic magnon at the zone center, see \cite{suppl-mat}.

The shift of the transversal magnon to the upper region of the magnon dispersion in \arc ~ resembles the dispersion of out-of-plane polarized magnons in Ca$_2$RuO$_4$ see \cite{jain2017,kunkemoeller2017}. But in contrast to Ca$_2$RuO$_4$, \arc ~ does not 
exhibit any low-energy magnon polarized within the layers, which excludes
describing the magnetic Hamiltonian by a simple $XY$ model. The polarized data taken at (0,1,1) in the paramagnetic phase, see Fig.~\ref{fig:ins} (b) reveal the rapid suppression of the sharp magnon mode by the melting of long-range order but the anisotropic and thus bond-directional character of the still strong and broad signal persists.

Measurements at other antiferromagnetic Bragg peaks are presented in Fig.~\ref{fig:otherafm}. There always is a sharp magnetic feature that is essentially  ${\bf z}_n$ polarized (we recall that ${\bf z}_n$ is always parallel to orthorhombic ${\bf a}$) confirming our interpretation for (0,1,1). While the amplitude is rapidly suppressed with $Q_L$, there is only little dispersion, which will be further discussed below. The enhancement of SFx intensity at the end of the scan for (0,1,2) and further polarized studies on a thermal spectrometer \cite{a-rcl_in20} as well as unpolarized experiments \cite{banerjee2018} suggest that a transversal zone-center magnon is situated around 4.5\,meV.

\begin{figure}
	\includegraphics[width=0.99 \columnwidth]{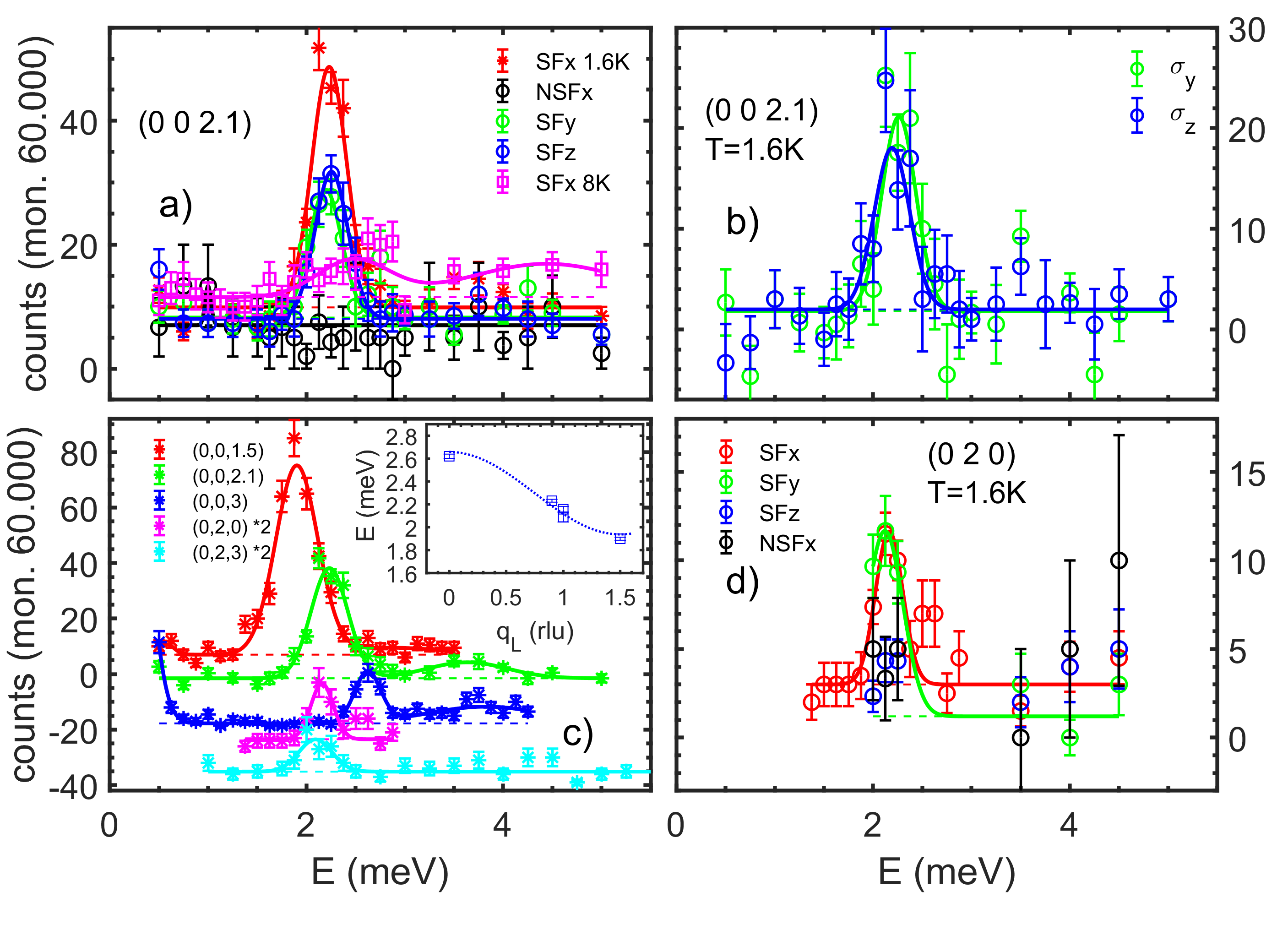}
	\caption{ {Polarized neutron scattering data at the 2D ferromagnetic zone centers (0,0,$Q_L$) and (0,2,$Q_L$) }}
	\label{fig:fm}
\end{figure}

Figure~\ref{fig:fm} presents the polarized data taken at the ferromagnetic scattering vectors. Panels 
(a) and (b) present the SF channels obtained at (0,0,2.1) and the resulting magnetic components, respectively. 
Again there is a sharp signal that agrees with unpolarized measurements \cite{balz2019} but this signal is isotropic.
For the magnetic response with scattering vector along the vertical axis of the trigonal material no anisotropy can be expected. 
The polarization analysis in the paramagnetic state confirms that the persisting scattering, which
is broad in energy and which is taken as a signature of the Kitaev coupling, indeed is magnetic. The ferromagnetic signal was studied at different $Q_L$ values and at (0,2,0) and (0,2,3) i.e. with a finite in-plane component. At (0,2,0) the usual polarization analysis could be performed revealing
a strongly anisotropic response. The quasi-ferromagnetic fluctuations sensed at (0,2,0) are essentially polarized along  ${\bf z}_n$ which corresponds to the in-plane direction vertical to the scattering vector, or
in other words the out-of-plane quasiferromagnetic fluctuations are strongly suppressed. This disagrees with the usual interpretation that this ferromagnetic response is predominantly induced by the ferromagnetic Kitaev interaction that implies an isotropic response.

The ferromagnetic response appearing at the zone center exhibits a pronounced $Q_L$ dispersion in the ordered state, which perfectly agrees with unpolarized earlier neutron measurements \cite{balz2019}. In order to compare the (0,0,$Q_L$) results with those
at (0,2,0) and (0,2,3) one must consider the non-conventional centring translations which result in extinction rules for Bragg positions. (0,2,0) is therefore not a Bragg point of the nuclear lattice but  (0,2,-1) and (0,2,2) are, therefore (0,2,0) and (0,2,3) correspond to ${\bf q}$=(0,0,1) and indeed the 
combined $Q_L$ dispersion is smooth. The ${\bf c}$ dispersion has been analyzed by Janssen et al. introducing 
additional inter-layer coupling terms \cite{janssen2020}. The shortest inter-layer distance was considered and
a rather large value of this interaction was introduced in order to describe the sizable ${\bf c}$ dispersion of the ferromagnetic signal. 
However this parameter implies an even stronger dispersion of the
antiferromagnetic fluctuations \cite{suppl-mat} for which we only find a weak dispersion, see Fig. 3 d). 
Describing the ${\bf c}$-axis dispersion
with the next-nearest inter-layer distances seems more appropriate. As it is explained in the supplemental information \cite{suppl-mat}  choosing J$_p'$=0.04\,meV we can well reproduce the ${\bf c}$-axis dispersion of the ferromagnetic and
of the antiferromagnetic excitations with a rather small parameter. This holds for the
three different models analyzed \cite{winter2017,liu2022,maksimov2020}. 
\arc ~ clearly shows finite vertical dispersion, but its dominant 2D character is confirmed by the weakness of the interlayer
interaction that is about two orders of magnitude weaker than the strongest intra-layer parameters.

In conclusion polarized neutron scattering experiments can further elucidate the magnetic interaction
of the most promising Kitaev candidate material \arc . The polarization of inelastic low-energy response is just opposite to a simple Heisenberg model confirming the anisotropic and bond-directional
character of the magnetic interaction. The usual transversal magnon modes are absent at low energies and are pushed to higher energies by the bond-directional anisotropy. However, other findings disagree with a simple or strongly dominant Kitaev interaction as the response at the 2D ferromagnetic scattering vectors is anisotropic with suppressed vertically polarized excitations.

\begin{acknowledgments}

We acknowledge support by the Deutsche Forschungsgemeinschaft (DFG, German Research Foundation) - Project number 277146847 - CRC 1238, project B04. 
\end{acknowledgments}


\begin{thebibliography}{51}%
	\makeatletter
	\providecommand \@ifxundefined [1]{%
		\@ifx{#1\undefined}
	}%
	\providecommand \@ifnum [1]{%
		\ifnum #1\expandafter \@firstoftwo
		\else \expandafter \@secondoftwo
		\fi
	}%
	\providecommand \@ifx [1]{%
		\ifx #1\expandafter \@firstoftwo
		\else \expandafter \@secondoftwo
		\fi
	}%
	\providecommand \natexlab [1]{#1}%
	\providecommand \enquote  [1]{``#1''}%
	\providecommand \bibnamefont  [1]{#1}%
	\providecommand \bibfnamefont [1]{#1}%
	\providecommand \citenamefont [1]{#1}%
	\providecommand \href@noop [0]{\@secondoftwo}%
	\providecommand \href [0]{\begingroup \@sanitize@url \@href}%
	\providecommand \@href[1]{\@@startlink{#1}\@@href}%
	\providecommand \@@href[1]{\endgroup#1\@@endlink}%
	\providecommand \@sanitize@url [0]{\catcode `\\12\catcode `\$12\catcode
		`\&12\catcode `\#12\catcode `\^12\catcode `\_12\catcode `\%12\relax}%
	\providecommand \@@startlink[1]{}%
	\providecommand \@@endlink[0]{}%
	\providecommand \url  [0]{\begingroup\@sanitize@url \@url }%
	\providecommand \@url [1]{\endgroup\@href {#1}{\urlprefix }}%
	\providecommand \urlprefix  [0]{URL }%
	\providecommand \Eprint [0]{\href }%
	\providecommand \doibase [0]{https://doi.org/}%
	\providecommand \selectlanguage [0]{\@gobble}%
	\providecommand \bibinfo  [0]{\@secondoftwo}%
	\providecommand \bibfield  [0]{\@secondoftwo}%
	\providecommand \translation [1]{[#1]}%
	\providecommand \BibitemOpen [0]{}%
	\providecommand \bibitemStop [0]{}%
	\providecommand \bibitemNoStop [0]{.\EOS\space}%
	\providecommand \EOS [0]{\spacefactor3000\relax}%
	\providecommand \BibitemShut  [1]{\csname bibitem#1\endcsname}%
	\let\auto@bib@innerbib\@empty
	\bibitem [{\citenamefont {Kitaev}(2006)}]{kitaev2006}%
	\BibitemOpen
	\bibfield  {author} {\bibinfo {author} {\bibfnamefont {A.}~\bibnamefont
			{Kitaev}},\ }\bibfield  {title} {\bibinfo {title} {Anyons in an exactly
			solved model and beyond},\ }\href
	{https://doi.org/https://doi.org/10.1016/j.aop.2005.10.005} {\bibfield
		{journal} {\bibinfo  {journal} {Annals of Physics}\ }\textbf {\bibinfo
			{volume} {321}},\ \bibinfo {pages} {2} (\bibinfo {year} {2006})}\BibitemShut
	{NoStop}%
	\bibitem [{\citenamefont {Trebst}\ and\ \citenamefont
		{Hickey}(2022)}]{trebst2022}%
	\BibitemOpen
	\bibfield  {author} {\bibinfo {author} {\bibfnamefont {S.}~\bibnamefont
			{Trebst}}\ and\ \bibinfo {author} {\bibfnamefont {C.}~\bibnamefont
			{Hickey}},\ }\bibfield  {title} {\bibinfo {title} {{Kitaev materials}},\
	}\href {https://doi.org/https://doi.org/10.1016/j.physrep.2021.11.003}
	{\bibfield  {journal} {\bibinfo  {journal} {Physics Reports}\ }\textbf
		{\bibinfo {volume} {950}},\ \bibinfo {pages} {1} (\bibinfo {year}
		{2022})}\BibitemShut {NoStop}%
	\bibitem [{\citenamefont {Takagi}\ \emph {et~al.}(2019)\citenamefont {Takagi},
		\citenamefont {Takayama}, \citenamefont {Jackeli}, \citenamefont
		{Khaliullin},\ and\ \citenamefont {Nagler}}]{takagi2019}%
	\BibitemOpen
	\bibfield  {author} {\bibinfo {author} {\bibfnamefont {H.}~\bibnamefont
			{Takagi}}, \bibinfo {author} {\bibfnamefont {T.}~\bibnamefont {Takayama}},
		\bibinfo {author} {\bibfnamefont {G.}~\bibnamefont {Jackeli}}, \bibinfo
		{author} {\bibfnamefont {G.}~\bibnamefont {Khaliullin}},\ and\ \bibinfo
		{author} {\bibfnamefont {S.~E.}\ \bibnamefont {Nagler}},\ }\bibfield  {title}
	{\bibinfo {title} {{Concept and realization of Kitaev quantum spin
				liquids}},\ }\href@noop {} {\bibfield  {journal} {\bibinfo  {journal} {Nature
				Reviews Physics}\ }\textbf {\bibinfo {volume} {1}},\ \bibinfo {pages} {264}
		(\bibinfo {year} {2019})}\BibitemShut {NoStop}%
	\bibitem [{\citenamefont {Winter}\ \emph
		{et~al.}(2017{\natexlab{a}})\citenamefont {Winter}, \citenamefont {Tsirlin},
		\citenamefont {Daghofer}, \citenamefont {van~den Brink}, \citenamefont
		{Singh}, \citenamefont {Gegenwart},\ and\ \citenamefont
		{Valent\'i}}]{winter2017b}%
	\BibitemOpen
	\bibfield  {author} {\bibinfo {author} {\bibfnamefont {S.~M.}\ \bibnamefont
			{Winter}}, \bibinfo {author} {\bibfnamefont {A.~A.}\ \bibnamefont {Tsirlin}},
		\bibinfo {author} {\bibfnamefont {M.}~\bibnamefont {Daghofer}}, \bibinfo
		{author} {\bibfnamefont {J.}~\bibnamefont {van~den Brink}}, \bibinfo {author}
		{\bibfnamefont {Y.}~\bibnamefont {Singh}}, \bibinfo {author} {\bibfnamefont
			{P.}~\bibnamefont {Gegenwart}},\ and\ \bibinfo {author} {\bibfnamefont
			{R.}~\bibnamefont {Valent\'i}},\ }\bibfield  {title} {\bibinfo {title}
		{{Models and materials for generalized Kitaev magnetism}},\ }\href
	{https://doi.org/10.1088/1361-648X/aa8cf5} {\bibfield  {journal} {\bibinfo
			{journal} {J. Phys.: Cond. Matt.}\ }\textbf {\bibinfo {volume} {29}},\
		\bibinfo {pages} {493002} (\bibinfo {year} {2017}{\natexlab{a}})}\BibitemShut
	{NoStop}%
	\bibitem [{\citenamefont {Broholm}\ \emph {et~al.}(2020)\citenamefont
		{Broholm}, \citenamefont {Cava}, \citenamefont {Kivelson}, \citenamefont
		{Nocera}, \citenamefont {Norman},\ and\ \citenamefont
		{Senthil}}]{broholm2020}%
	\BibitemOpen
	\bibfield  {author} {\bibinfo {author} {\bibfnamefont {C.}~\bibnamefont
			{Broholm}}, \bibinfo {author} {\bibfnamefont {R.~J.}\ \bibnamefont {Cava}},
		\bibinfo {author} {\bibfnamefont {S.~A.}\ \bibnamefont {Kivelson}}, \bibinfo
		{author} {\bibfnamefont {D.~G.}\ \bibnamefont {Nocera}}, \bibinfo {author}
		{\bibfnamefont {M.~R.}\ \bibnamefont {Norman}},\ and\ \bibinfo {author}
		{\bibfnamefont {T.}~\bibnamefont {Senthil}},\ }\bibfield  {title} {\bibinfo
		{title} {Quantum spin liquids},\ }\href
	{https://doi.org/10.1126/science.aay0668} {\bibfield  {journal} {\bibinfo
			{journal} {Science}\ }\textbf {\bibinfo {volume} {367}},\ \bibinfo {pages}
		{eaay0668} (\bibinfo {year} {2020})}\BibitemShut {NoStop}%
	\bibitem [{\citenamefont {Singh}\ and\ \citenamefont
		{Gegenwart}(2010)}]{singh2010}%
	\BibitemOpen
	\bibfield  {author} {\bibinfo {author} {\bibfnamefont {Y.}~\bibnamefont
			{Singh}}\ and\ \bibinfo {author} {\bibfnamefont {P.}~\bibnamefont
			{Gegenwart}},\ }\bibfield  {title} {\bibinfo {title} {{Antiferromagnetic Mott
				insulating state in single crystals of the honeycomb lattice material
				${\text{Na}}_{2}{\text{IrO}}_{3}$}},\ }\href
	{https://doi.org/10.1103/PhysRevB.82.064412} {\bibfield  {journal} {\bibinfo
			{journal} {Phys. Rev. B}\ }\textbf {\bibinfo {volume} {82}},\ \bibinfo
		{pages} {064412} (\bibinfo {year} {2010})}\BibitemShut {NoStop}%
	\bibitem [{\citenamefont {Chaloupka}\ \emph {et~al.}(2010)\citenamefont
		{Chaloupka}, \citenamefont {Jackeli},\ and\ \citenamefont
		{Khaliullin}}]{chaloupka2010}%
	\BibitemOpen
	\bibfield  {author} {\bibinfo {author} {\bibfnamefont {J.}~\bibnamefont
			{Chaloupka}}, \bibinfo {author} {\bibfnamefont {G.}~\bibnamefont {Jackeli}},\
		and\ \bibinfo {author} {\bibfnamefont {G.}~\bibnamefont {Khaliullin}},\
	}\bibfield  {title} {\bibinfo {title} {{Kitaev-Heisenberg Model on a
				Honeycomb Lattice: Possible Exotic Phases in Iridium Oxides
				${A}_{2}{\mathrm{IrO}}_{3}$}},\ }\href
	{https://doi.org/10.1103/PhysRevLett.105.027204} {\bibfield  {journal}
		{\bibinfo  {journal} {Phys. Rev. Lett.}\ }\textbf {\bibinfo {volume} {105}},\
		\bibinfo {pages} {027204} (\bibinfo {year} {2010})}\BibitemShut {NoStop}%
	\bibitem [{\citenamefont {Jackeli}\ and\ \citenamefont
		{Khaliullin}(2009)}]{jackeli2009}%
	\BibitemOpen
	\bibfield  {author} {\bibinfo {author} {\bibfnamefont {G.}~\bibnamefont
			{Jackeli}}\ and\ \bibinfo {author} {\bibfnamefont {G.}~\bibnamefont
			{Khaliullin}},\ }\bibfield  {title} {\bibinfo {title} {{Mott Insulators in
				the Strong Spin-Orbit Coupling Limit: From Heisenberg to a Quantum Compass
				and Kitaev Models}},\ }\href {https://doi.org/10.1103/PhysRevLett.102.017205}
	{\bibfield  {journal} {\bibinfo  {journal} {Phys. Rev. Lett.}\ }\textbf
		{\bibinfo {volume} {102}},\ \bibinfo {pages} {017205} (\bibinfo {year}
		{2009})}\BibitemShut {NoStop}%
	\bibitem [{\citenamefont {Plumb}\ \emph {et~al.}(2014)\citenamefont {Plumb},
		\citenamefont {Clancy}, \citenamefont {Sandilands}, \citenamefont {Shankar},
		\citenamefont {Hu}, \citenamefont {Burch}, \citenamefont {Kee},\ and\
		\citenamefont {Kim}}]{plumb2014}%
	\BibitemOpen
	\bibfield  {author} {\bibinfo {author} {\bibfnamefont {K.~W.}\ \bibnamefont
			{Plumb}}, \bibinfo {author} {\bibfnamefont {J.~P.}\ \bibnamefont {Clancy}},
		\bibinfo {author} {\bibfnamefont {L.~J.}\ \bibnamefont {Sandilands}},
		\bibinfo {author} {\bibfnamefont {V.~V.}\ \bibnamefont {Shankar}}, \bibinfo
		{author} {\bibfnamefont {Y.~F.}\ \bibnamefont {Hu}}, \bibinfo {author}
		{\bibfnamefont {K.~S.}\ \bibnamefont {Burch}}, \bibinfo {author}
		{\bibfnamefont {H.-Y.}\ \bibnamefont {Kee}},\ and\ \bibinfo {author}
		{\bibfnamefont {Y.-J.}\ \bibnamefont {Kim}},\ }\bibfield  {title} {\bibinfo
		{title} {{$\ensuremath{\alpha}$-${\mathrm{RuCl}}_{3}$: A spin-orbit assisted
				Mott insulator on a honeycomb lattice}},\ }\href
	{https://doi.org/10.1103/PhysRevB.90.041112} {\bibfield  {journal} {\bibinfo
			{journal} {Phys. Rev. B}\ }\textbf {\bibinfo {volume} {90}},\ \bibinfo
		{pages} {041112} (\bibinfo {year} {2014})}\BibitemShut {NoStop}%
	\bibitem [{\citenamefont {Kasahara}\ \emph {et~al.}(2018)\citenamefont
		{Kasahara}, \citenamefont {Ohnishi}, \citenamefont {Mizukami}, \citenamefont
		{Tanaka}, \citenamefont {Ma}, \citenamefont {Sugii}, \citenamefont {Kurita},
		\citenamefont {Tanaka}, \citenamefont {Nasu}, \citenamefont {Motome},
		\citenamefont {Shibauchi},\ and\ \citenamefont {Matsuda}}]{kasahara2018}%
	\BibitemOpen
	\bibfield  {author} {\bibinfo {author} {\bibfnamefont {Y.}~\bibnamefont
			{Kasahara}}, \bibinfo {author} {\bibfnamefont {T.}~\bibnamefont {Ohnishi}},
		\bibinfo {author} {\bibfnamefont {Y.}~\bibnamefont {Mizukami}}, \bibinfo
		{author} {\bibfnamefont {O.}~\bibnamefont {Tanaka}}, \bibinfo {author}
		{\bibfnamefont {S.}~\bibnamefont {Ma}}, \bibinfo {author} {\bibfnamefont
			{K.}~\bibnamefont {Sugii}}, \bibinfo {author} {\bibfnamefont
			{N.}~\bibnamefont {Kurita}}, \bibinfo {author} {\bibfnamefont
			{H.}~\bibnamefont {Tanaka}}, \bibinfo {author} {\bibfnamefont
			{J.}~\bibnamefont {Nasu}}, \bibinfo {author} {\bibfnamefont {Y.}~\bibnamefont
			{Motome}}, \bibinfo {author} {\bibfnamefont {T.}~\bibnamefont {Shibauchi}},\
		and\ \bibinfo {author} {\bibfnamefont {Y.}~\bibnamefont {Matsuda}},\
	}\bibfield  {title} {\bibinfo {title} {{Majorana quantization and
				half-integer thermal quantum Hall effect in a Kitaev spin liquid}},\ }\href
	{https://doi.org/10.1038/s41586-018-0274-0} {\bibfield  {journal} {\bibinfo
			{journal} {Nature}\ }\textbf {\bibinfo {volume} {559}},\ \bibinfo {pages}
		{227} (\bibinfo {year} {2018})}\BibitemShut {NoStop}%
	\bibitem [{\citenamefont {Bruin}\ \emph {et~al.}(2022)\citenamefont {Bruin},
		\citenamefont {Claus}, \citenamefont {Matsumoto}, \citenamefont {Kurita},
		\citenamefont {Tanaka},\ and\ \citenamefont {Takagi}}]{bruin2022}%
	\BibitemOpen
	\bibfield  {author} {\bibinfo {author} {\bibfnamefont {J.}~\bibnamefont
			{Bruin}}, \bibinfo {author} {\bibfnamefont {R.}~\bibnamefont {Claus}},
		\bibinfo {author} {\bibfnamefont {Y.}~\bibnamefont {Matsumoto}}, \bibinfo
		{author} {\bibfnamefont {N.}~\bibnamefont {Kurita}}, \bibinfo {author}
		{\bibfnamefont {H.}~\bibnamefont {Tanaka}},\ and\ \bibinfo {author}
		{\bibfnamefont {H.}~\bibnamefont {Takagi}},\ }\bibfield  {title} {\bibinfo
		{title} {{Robustness of the thermal Hall effect close to half-quantization in
				$\alpha$-RuCl$_3$}},\ }\href@noop {} {\bibfield  {journal} {\bibinfo
			{journal} {Nature Physics}\ }\textbf {\bibinfo {volume} {18}},\ \bibinfo
		{pages} {401} (\bibinfo {year} {2022})}\BibitemShut {NoStop}%
	\bibitem [{\citenamefont {Czajka}\ \emph {et~al.}(2021)\citenamefont {Czajka},
		\citenamefont {Gao}, \citenamefont {Hirschberger}, \citenamefont
		{Lampen-Kelley}, \citenamefont {Banerjee}, \citenamefont {Yan}, \citenamefont
		{Mandrus}, \citenamefont {Nagler},\ and\ \citenamefont {Ong}}]{czajka2021}%
	\BibitemOpen
	\bibfield  {author} {\bibinfo {author} {\bibfnamefont {P.}~\bibnamefont
			{Czajka}}, \bibinfo {author} {\bibfnamefont {T.}~\bibnamefont {Gao}},
		\bibinfo {author} {\bibfnamefont {M.}~\bibnamefont {Hirschberger}}, \bibinfo
		{author} {\bibfnamefont {P.}~\bibnamefont {Lampen-Kelley}}, \bibinfo {author}
		{\bibfnamefont {A.}~\bibnamefont {Banerjee}}, \bibinfo {author}
		{\bibfnamefont {J.}~\bibnamefont {Yan}}, \bibinfo {author} {\bibfnamefont
			{D.~G.}\ \bibnamefont {Mandrus}}, \bibinfo {author} {\bibfnamefont {S.~E.}\
			\bibnamefont {Nagler}},\ and\ \bibinfo {author} {\bibfnamefont {N.~P.}\
			\bibnamefont {Ong}},\ }\bibfield  {title} {\bibinfo {title} {{Oscillations of
				the thermal conductivity in the spin-liquid state of $\alpha$-RuCl$_3$}},\
	}\href {https://doi.org/10.1038/s41567-021-01243-x} {\bibfield  {journal}
		{\bibinfo  {journal} {Nature Physics}\ }\textbf {\bibinfo {volume} {17}},\
		\bibinfo {pages} {915} (\bibinfo {year} {2021})}\BibitemShut {NoStop}%
	\bibitem [{\citenamefont {Johnson}\ \emph {et~al.}(2015)\citenamefont
		{Johnson}, \citenamefont {Williams}, \citenamefont {Haghighirad},
		\citenamefont {Singleton}, \citenamefont {Zapf}, \citenamefont {Manuel},
		\citenamefont {Mazin}, \citenamefont {Li}, \citenamefont {Jeschke},
		\citenamefont {Valent\'{\i}},\ and\ \citenamefont {Coldea}}]{johnson2015}%
	\BibitemOpen
	\bibfield  {author} {\bibinfo {author} {\bibfnamefont {R.~D.}\ \bibnamefont
			{Johnson}}, \bibinfo {author} {\bibfnamefont {S.~C.}\ \bibnamefont
			{Williams}}, \bibinfo {author} {\bibfnamefont {A.~A.}\ \bibnamefont
			{Haghighirad}}, \bibinfo {author} {\bibfnamefont {J.}~\bibnamefont
			{Singleton}}, \bibinfo {author} {\bibfnamefont {V.}~\bibnamefont {Zapf}},
		\bibinfo {author} {\bibfnamefont {P.}~\bibnamefont {Manuel}}, \bibinfo
		{author} {\bibfnamefont {I.~I.}\ \bibnamefont {Mazin}}, \bibinfo {author}
		{\bibfnamefont {Y.}~\bibnamefont {Li}}, \bibinfo {author} {\bibfnamefont
			{H.~O.}\ \bibnamefont {Jeschke}}, \bibinfo {author} {\bibfnamefont
			{R.}~\bibnamefont {Valent\'{\i}}},\ and\ \bibinfo {author} {\bibfnamefont
			{R.}~\bibnamefont {Coldea}},\ }\bibfield  {title} {\bibinfo {title}
		{{Monoclinic crystal structure of
				$\ensuremath{\alpha}\ensuremath{-}{\mathrm{RuCl}}_{3}$ and the zigzag
				antiferromagnetic ground state}},\ }\href
	{https://doi.org/10.1103/PhysRevB.92.235119} {\bibfield  {journal} {\bibinfo
			{journal} {Phys. Rev. B}\ }\textbf {\bibinfo {volume} {92}},\ \bibinfo
		{pages} {235119} (\bibinfo {year} {2015})}\BibitemShut {NoStop}%
	\bibitem [{\citenamefont {Cao}\ \emph {et~al.}(2016)\citenamefont {Cao},
		\citenamefont {Banerjee}, \citenamefont {Yan}, \citenamefont {Bridges},
		\citenamefont {Lumsden}, \citenamefont {Mandrus}, \citenamefont {Tennant},
		\citenamefont {Chakoumakos},\ and\ \citenamefont {Nagler}}]{cao2016}%
	\BibitemOpen
	\bibfield  {author} {\bibinfo {author} {\bibfnamefont {H.~B.}\ \bibnamefont
			{Cao}}, \bibinfo {author} {\bibfnamefont {A.}~\bibnamefont {Banerjee}},
		\bibinfo {author} {\bibfnamefont {J.-Q.}\ \bibnamefont {Yan}}, \bibinfo
		{author} {\bibfnamefont {C.~A.}\ \bibnamefont {Bridges}}, \bibinfo {author}
		{\bibfnamefont {M.~D.}\ \bibnamefont {Lumsden}}, \bibinfo {author}
		{\bibfnamefont {D.~G.}\ \bibnamefont {Mandrus}}, \bibinfo {author}
		{\bibfnamefont {D.~A.}\ \bibnamefont {Tennant}}, \bibinfo {author}
		{\bibfnamefont {B.~C.}\ \bibnamefont {Chakoumakos}},\ and\ \bibinfo {author}
		{\bibfnamefont {S.~E.}\ \bibnamefont {Nagler}},\ }\bibfield  {title}
	{\bibinfo {title} {{Low-temperature crystal and magnetic structure of
				$\ensuremath{\alpha}\ensuremath{-}{\mathrm{RuCl}}_{3}$}},\ }\href
	{https://doi.org/10.1103/PhysRevB.93.134423} {\bibfield  {journal} {\bibinfo
			{journal} {Phys. Rev. B}\ }\textbf {\bibinfo {volume} {93}},\ \bibinfo
		{pages} {134423} (\bibinfo {year} {2016})}\BibitemShut {NoStop}%
	\bibitem [{\citenamefont {Mazin}\ \emph {et~al.}(2012)\citenamefont {Mazin},
		\citenamefont {Jeschke}, \citenamefont {Foyevtsova}, \citenamefont
		{Valent\'{\i}},\ and\ \citenamefont {Khomskii}}]{mazin2012}%
	\BibitemOpen
	\bibfield  {author} {\bibinfo {author} {\bibfnamefont {I.~I.}\ \bibnamefont
			{Mazin}}, \bibinfo {author} {\bibfnamefont {H.~O.}\ \bibnamefont {Jeschke}},
		\bibinfo {author} {\bibfnamefont {K.}~\bibnamefont {Foyevtsova}}, \bibinfo
		{author} {\bibfnamefont {R.}~\bibnamefont {Valent\'{\i}}},\ and\ \bibinfo
		{author} {\bibfnamefont {D.~I.}\ \bibnamefont {Khomskii}},\ }\bibfield
	{title} {\bibinfo {title} {{${\mathrm{Na}}_{2}{\mathrm{IrO}}_{3}$ as a
				Molecular Orbital Crystal}},\ }\href
	{https://doi.org/10.1103/PhysRevLett.109.197201} {\bibfield  {journal}
		{\bibinfo  {journal} {Phys. Rev. Lett.}\ }\textbf {\bibinfo {volume} {109}},\
		\bibinfo {pages} {197201} (\bibinfo {year} {2012})}\BibitemShut {NoStop}%
	\bibitem [{\citenamefont {Foyevtsova}\ \emph {et~al.}(2013)\citenamefont
		{Foyevtsova}, \citenamefont {Jeschke}, \citenamefont {Mazin}, \citenamefont
		{Khomskii},\ and\ \citenamefont {Valent\'{\i}}}]{foyevtsova2013}%
	\BibitemOpen
	\bibfield  {author} {\bibinfo {author} {\bibfnamefont {K.}~\bibnamefont
			{Foyevtsova}}, \bibinfo {author} {\bibfnamefont {H.~O.}\ \bibnamefont
			{Jeschke}}, \bibinfo {author} {\bibfnamefont {I.~I.}\ \bibnamefont {Mazin}},
		\bibinfo {author} {\bibfnamefont {D.~I.}\ \bibnamefont {Khomskii}},\ and\
		\bibinfo {author} {\bibfnamefont {R.}~\bibnamefont {Valent\'{\i}}},\
	}\bibfield  {title} {\bibinfo {title} {{Ab initio analysis of the
				tight-binding parameters and magnetic interactions in
				Na${}_{2}$IrO${}_{3}$}},\ }\href {https://doi.org/10.1103/PhysRevB.88.035107}
	{\bibfield  {journal} {\bibinfo  {journal} {Phys. Rev. B}\ }\textbf {\bibinfo
			{volume} {88}},\ \bibinfo {pages} {035107} (\bibinfo {year}
		{2013})}\BibitemShut {NoStop}%
	\bibitem [{\citenamefont {Winter}\ \emph
		{et~al.}(2017{\natexlab{b}})\citenamefont {Winter}, \citenamefont {Riedl},
		\citenamefont {Maksimov}, \citenamefont {Chernyshev}, \citenamefont
		{Honecker},\ and\ \citenamefont {Valent{\'i}}}]{winter2017}%
	\BibitemOpen
	\bibfield  {author} {\bibinfo {author} {\bibfnamefont {S.~M.}\ \bibnamefont
			{Winter}}, \bibinfo {author} {\bibfnamefont {K.}~\bibnamefont {Riedl}},
		\bibinfo {author} {\bibfnamefont {P.~A.}\ \bibnamefont {Maksimov}}, \bibinfo
		{author} {\bibfnamefont {A.~L.}\ \bibnamefont {Chernyshev}}, \bibinfo
		{author} {\bibfnamefont {A.}~\bibnamefont {Honecker}},\ and\ \bibinfo
		{author} {\bibfnamefont {R.}~\bibnamefont {Valent{\'i}}},\ }\bibfield
	{title} {\bibinfo {title} {Breakdown of magnons in a strongly spin-orbital
			coupled magnet},\ }\href {https://doi.org/10.1038/s41467-017-01177-0}
	{\bibfield  {journal} {\bibinfo  {journal} {Nature Communications}\ }\textbf
		{\bibinfo {volume} {8}},\ \bibinfo {pages} {1152} (\bibinfo {year}
		{2017}{\natexlab{b}})}\BibitemShut {NoStop}%
	\bibitem [{\citenamefont {Liu}\ \emph {et~al.}(2022)\citenamefont {Liu},
		\citenamefont {Chaloupka},\ and\ \citenamefont {Khaliullin}}]{liu2022}%
	\BibitemOpen
	\bibfield  {author} {\bibinfo {author} {\bibfnamefont {H.}~\bibnamefont
			{Liu}}, \bibinfo {author} {\bibfnamefont {J.}~\bibnamefont {Chaloupka}},\
		and\ \bibinfo {author} {\bibfnamefont {G.}~\bibnamefont {Khaliullin}},\
	}\bibfield  {title} {\bibinfo {title} {{Exchange interactions in ${d}^{5}$
				Kitaev materials: From Na$_2$IrO$_{3}$ to $\alpha$-RuCl$_3$}},\ }\href
	{https://doi.org/10.1103/PhysRevB.105.214411} {\bibfield  {journal} {\bibinfo
			{journal} {Phys. Rev. B}\ }\textbf {\bibinfo {volume} {105}},\ \bibinfo
		{pages} {214411} (\bibinfo {year} {2022})}\BibitemShut {NoStop}%
	\bibitem [{\citenamefont {Banerjee}\ \emph {et~al.}(2016)\citenamefont
		{Banerjee}, \citenamefont {Bridges}, \citenamefont {Yan}, \citenamefont
		{Aczel}, \citenamefont {Li}, \citenamefont {Stone}, \citenamefont {Granroth},
		\citenamefont {Lumsden}, \citenamefont {Yiu}, \citenamefont {Knolle},
		\citenamefont {Bhattacharjee}, \citenamefont {Kovrizhin}, \citenamefont
		{Moessner}, \citenamefont {Tennant}, \citenamefont {Mandrus},\ and\
		\citenamefont {Nagler}}]{banerjee2016}%
	\BibitemOpen
	\bibfield  {author} {\bibinfo {author} {\bibfnamefont {A.}~\bibnamefont
			{Banerjee}}, \bibinfo {author} {\bibfnamefont {C.~A.}\ \bibnamefont
			{Bridges}}, \bibinfo {author} {\bibfnamefont {J.-Q.}\ \bibnamefont {Yan}},
		\bibinfo {author} {\bibfnamefont {A.~A.}\ \bibnamefont {Aczel}}, \bibinfo
		{author} {\bibfnamefont {L.}~\bibnamefont {Li}}, \bibinfo {author}
		{\bibfnamefont {M.~B.}\ \bibnamefont {Stone}}, \bibinfo {author}
		{\bibfnamefont {G.~E.}\ \bibnamefont {Granroth}}, \bibinfo {author}
		{\bibfnamefont {M.~D.}\ \bibnamefont {Lumsden}}, \bibinfo {author}
		{\bibfnamefont {Y.}~\bibnamefont {Yiu}}, \bibinfo {author} {\bibfnamefont
			{J.}~\bibnamefont {Knolle}}, \bibinfo {author} {\bibfnamefont
			{S.}~\bibnamefont {Bhattacharjee}}, \bibinfo {author} {\bibfnamefont {D.~L.}\
			\bibnamefont {Kovrizhin}}, \bibinfo {author} {\bibfnamefont {R.}~\bibnamefont
			{Moessner}}, \bibinfo {author} {\bibfnamefont {D.~A.}\ \bibnamefont
			{Tennant}}, \bibinfo {author} {\bibfnamefont {D.~G.}\ \bibnamefont
			{Mandrus}},\ and\ \bibinfo {author} {\bibfnamefont {S.~E.}\ \bibnamefont
			{Nagler}},\ }\bibfield  {title} {\bibinfo {title} {{Proximate Kitaev quantum
				spin liquid behaviour in a honeycomb magnet}},\ }\href
	{https://doi.org/10.1038/nmat4604} {\bibfield  {journal} {\bibinfo  {journal}
			{Nature Materials}\ }\textbf {\bibinfo {volume} {15}},\ \bibinfo {pages}
		{733} (\bibinfo {year} {2016})}\BibitemShut {NoStop}%
	\bibitem [{\citenamefont {Banerjee}\ \emph {et~al.}(2017)\citenamefont
		{Banerjee}, \citenamefont {Yan}, \citenamefont {Knolle}, \citenamefont
		{Bridges}, \citenamefont {Stone}, \citenamefont {Lumsden}, \citenamefont
		{Mandrus}, \citenamefont {Tennant}, \citenamefont {Moessner},\ and\
		\citenamefont {Nagler}}]{banerjee2017}%
	\BibitemOpen
	\bibfield  {author} {\bibinfo {author} {\bibfnamefont {A.}~\bibnamefont
			{Banerjee}}, \bibinfo {author} {\bibfnamefont {J.}~\bibnamefont {Yan}},
		\bibinfo {author} {\bibfnamefont {J.}~\bibnamefont {Knolle}}, \bibinfo
		{author} {\bibfnamefont {C.~A.}\ \bibnamefont {Bridges}}, \bibinfo {author}
		{\bibfnamefont {M.~B.}\ \bibnamefont {Stone}}, \bibinfo {author}
		{\bibfnamefont {M.~D.}\ \bibnamefont {Lumsden}}, \bibinfo {author}
		{\bibfnamefont {D.~G.}\ \bibnamefont {Mandrus}}, \bibinfo {author}
		{\bibfnamefont {D.~A.}\ \bibnamefont {Tennant}}, \bibinfo {author}
		{\bibfnamefont {R.}~\bibnamefont {Moessner}},\ and\ \bibinfo {author}
		{\bibfnamefont {S.~E.}\ \bibnamefont {Nagler}},\ }\bibfield  {title}
	{\bibinfo {title} {{Neutron scattering in the proximate quantum spin liquid
				$\alpha$-RuCl$_3$}},\ }\href {https://doi.org/10.1126/science.aah6015}
	{\bibfield  {journal} {\bibinfo  {journal} {Science}\ }\textbf {\bibinfo
			{volume} {356}},\ \bibinfo {pages} {1055} (\bibinfo {year}
		{2017})}\BibitemShut {NoStop}%
	\bibitem [{\citenamefont {Banerjee}\ \emph {et~al.}(2018)\citenamefont
		{Banerjee}, \citenamefont {Lampen-Kelley}, \citenamefont {Knolle},
		\citenamefont {Balz}, \citenamefont {Aczel}, \citenamefont {Winn},
		\citenamefont {Liu}, \citenamefont {Pajerowski}, \citenamefont {Yan},
		\citenamefont {Bridges}, \citenamefont {Savici}, \citenamefont {Chakoumakos},
		\citenamefont {Lumsden}, \citenamefont {Tennant}, \citenamefont {Moessner},
		\citenamefont {Mandrus},\ and\ \citenamefont {Nagler}}]{banerjee2018}%
	\BibitemOpen
	\bibfield  {author} {\bibinfo {author} {\bibfnamefont {A.}~\bibnamefont
			{Banerjee}}, \bibinfo {author} {\bibfnamefont {P.}~\bibnamefont
			{Lampen-Kelley}}, \bibinfo {author} {\bibfnamefont {J.}~\bibnamefont
			{Knolle}}, \bibinfo {author} {\bibfnamefont {C.}~\bibnamefont {Balz}},
		\bibinfo {author} {\bibfnamefont {A.~A.}\ \bibnamefont {Aczel}}, \bibinfo
		{author} {\bibfnamefont {B.}~\bibnamefont {Winn}}, \bibinfo {author}
		{\bibfnamefont {Y.}~\bibnamefont {Liu}}, \bibinfo {author} {\bibfnamefont
			{D.}~\bibnamefont {Pajerowski}}, \bibinfo {author} {\bibfnamefont
			{J.}~\bibnamefont {Yan}}, \bibinfo {author} {\bibfnamefont {C.~A.}\
			\bibnamefont {Bridges}}, \bibinfo {author} {\bibfnamefont {A.~T.}\
			\bibnamefont {Savici}}, \bibinfo {author} {\bibfnamefont {B.~C.}\
			\bibnamefont {Chakoumakos}}, \bibinfo {author} {\bibfnamefont {M.~D.}\
			\bibnamefont {Lumsden}}, \bibinfo {author} {\bibfnamefont {D.~A.}\
			\bibnamefont {Tennant}}, \bibinfo {author} {\bibfnamefont {R.}~\bibnamefont
			{Moessner}}, \bibinfo {author} {\bibfnamefont {D.~G.}\ \bibnamefont
			{Mandrus}},\ and\ \bibinfo {author} {\bibfnamefont {S.~E.}\ \bibnamefont
			{Nagler}},\ }\bibfield  {title} {\bibinfo {title} {{Excitations in the
				field-induced quantum spin liquid state of $\alpha$-RuCl$_3$}},\ }\href
	{https://doi.org/10.1038/s41535-018-0079-2} {\bibfield  {journal} {\bibinfo
			{journal} {npj Quantum Materials}\ }\textbf {\bibinfo {volume} {3}},\
		\bibinfo {pages} {8} (\bibinfo {year} {2018})}\BibitemShut {NoStop}%
	\bibitem [{\citenamefont {Balz}\ \emph {et~al.}(2019)\citenamefont {Balz},
		\citenamefont {Lampen-Kelley}, \citenamefont {Banerjee}, \citenamefont {Yan},
		\citenamefont {Lu}, \citenamefont {Hu}, \citenamefont {Yadav}, \citenamefont
		{Takano}, \citenamefont {Liu}, \citenamefont {Tennant}, \citenamefont
		{Lumsden}, \citenamefont {Mandrus},\ and\ \citenamefont {Nagler}}]{balz2019}%
	\BibitemOpen
	\bibfield  {author} {\bibinfo {author} {\bibfnamefont {C.}~\bibnamefont
			{Balz}}, \bibinfo {author} {\bibfnamefont {P.}~\bibnamefont {Lampen-Kelley}},
		\bibinfo {author} {\bibfnamefont {A.}~\bibnamefont {Banerjee}}, \bibinfo
		{author} {\bibfnamefont {J.}~\bibnamefont {Yan}}, \bibinfo {author}
		{\bibfnamefont {Z.}~\bibnamefont {Lu}}, \bibinfo {author} {\bibfnamefont
			{X.}~\bibnamefont {Hu}}, \bibinfo {author} {\bibfnamefont {S.~M.}\
			\bibnamefont {Yadav}}, \bibinfo {author} {\bibfnamefont {Y.}~\bibnamefont
			{Takano}}, \bibinfo {author} {\bibfnamefont {Y.}~\bibnamefont {Liu}},
		\bibinfo {author} {\bibfnamefont {D.~A.}\ \bibnamefont {Tennant}}, \bibinfo
		{author} {\bibfnamefont {M.~D.}\ \bibnamefont {Lumsden}}, \bibinfo {author}
		{\bibfnamefont {D.}~\bibnamefont {Mandrus}},\ and\ \bibinfo {author}
		{\bibfnamefont {S.~E.}\ \bibnamefont {Nagler}},\ }\bibfield  {title}
	{\bibinfo {title} {{Finite field regime for a quantum spin liquid in
				$\ensuremath{\alpha}\text{\ensuremath{-}}{\mathrm{RuCl}}_{3}$}},\ }\href
	{https://doi.org/10.1103/PhysRevB.100.060405} {\bibfield  {journal} {\bibinfo
			{journal} {Phys. Rev. B}\ }\textbf {\bibinfo {volume} {100}},\ \bibinfo
		{pages} {060405} (\bibinfo {year} {2019})}\BibitemShut {NoStop}%
	\bibitem [{\citenamefont {Ran}\ \emph {et~al.}(2017)\citenamefont {Ran},
		\citenamefont {Wang}, \citenamefont {Wang}, \citenamefont {Dong},
		\citenamefont {Ren}, \citenamefont {Bao}, \citenamefont {Li}, \citenamefont
		{Ma}, \citenamefont {Gan}, \citenamefont {Zhang} \emph {et~al.}}]{ran2017}%
	\BibitemOpen
	\bibfield  {author} {\bibinfo {author} {\bibfnamefont {K.}~\bibnamefont
			{Ran}}, \bibinfo {author} {\bibfnamefont {J.}~\bibnamefont {Wang}}, \bibinfo
		{author} {\bibfnamefont {W.}~\bibnamefont {Wang}}, \bibinfo {author}
		{\bibfnamefont {Z.-Y.}\ \bibnamefont {Dong}}, \bibinfo {author}
		{\bibfnamefont {X.}~\bibnamefont {Ren}}, \bibinfo {author} {\bibfnamefont
			{S.}~\bibnamefont {Bao}}, \bibinfo {author} {\bibfnamefont {S.}~\bibnamefont
			{Li}}, \bibinfo {author} {\bibfnamefont {Z.}~\bibnamefont {Ma}}, \bibinfo
		{author} {\bibfnamefont {Y.}~\bibnamefont {Gan}}, \bibinfo {author}
		{\bibfnamefont {Y.}~\bibnamefont {Zhang}}, \emph {et~al.},\ }\bibfield
	{title} {\bibinfo {title} {{Spin-Wave Excitations Evidencing the Kitaev
				Interaction in Single Crystalline $\alpha$-RuCl$_3$}},\ }\href@noop {}
	{\bibfield  {journal} {\bibinfo  {journal} {Phys. Rev. Lett.}\ }\textbf
		{\bibinfo {volume} {118}},\ \bibinfo {pages} {107203} (\bibinfo {year}
		{2017})}\BibitemShut {NoStop}%
	\bibitem [{\citenamefont {Ran}\ \emph {et~al.}(2022)\citenamefont {Ran},
		\citenamefont {Wang}, \citenamefont {Bao}, \citenamefont {Cai}, \citenamefont
		{Shangguan}, \citenamefont {Ma}, \citenamefont {Wang}, \citenamefont {Dong},
		\citenamefont {Čermák}, \citenamefont {Schneidewind}, \citenamefont {Meng},
		\citenamefont {Lu}, \citenamefont {Yu}, \citenamefont {Li},\ and\
		\citenamefont {Wen}}]{ran2022}%
	\BibitemOpen
	\bibfield  {author} {\bibinfo {author} {\bibfnamefont {K.}~\bibnamefont
			{Ran}}, \bibinfo {author} {\bibfnamefont {J.}~\bibnamefont {Wang}}, \bibinfo
		{author} {\bibfnamefont {S.}~\bibnamefont {Bao}}, \bibinfo {author}
		{\bibfnamefont {Z.}~\bibnamefont {Cai}}, \bibinfo {author} {\bibfnamefont
			{Y.}~\bibnamefont {Shangguan}}, \bibinfo {author} {\bibfnamefont
			{Z.}~\bibnamefont {Ma}}, \bibinfo {author} {\bibfnamefont {W.}~\bibnamefont
			{Wang}}, \bibinfo {author} {\bibfnamefont {Z.-Y.}\ \bibnamefont {Dong}},
		\bibinfo {author} {\bibfnamefont {P.}~\bibnamefont {Čermák}}, \bibinfo
		{author} {\bibfnamefont {A.}~\bibnamefont {Schneidewind}}, \bibinfo {author}
		{\bibfnamefont {S.}~\bibnamefont {Meng}}, \bibinfo {author} {\bibfnamefont
			{Z.}~\bibnamefont {Lu}}, \bibinfo {author} {\bibfnamefont {S.-L.}\
			\bibnamefont {Yu}}, \bibinfo {author} {\bibfnamefont {J.-X.}\ \bibnamefont
			{Li}},\ and\ \bibinfo {author} {\bibfnamefont {J.}~\bibnamefont {Wen}},\
	}\bibfield  {title} {\bibinfo {title} {{Evidence for Magnetic Fractional
				Excitations in a Kitaev Quantum-Spin-Liquid Candidate $\alpha$-RuCl$_3$}},\
	}\href {https://doi.org/10.1088/0256-307X/39/2/027501} {\bibfield  {journal}
		{\bibinfo  {journal} {Chinese Physics Letters}\ }\textbf {\bibinfo {volume}
			{39}},\ \bibinfo {pages} {027501} (\bibinfo {year} {2022})}\BibitemShut
	{NoStop}%
	\bibitem [{\citenamefont {Do}\ \emph {et~al.}(2017)\citenamefont {Do},
		\citenamefont {Park}, \citenamefont {Yoshitake}, \citenamefont {Nasu},
		\citenamefont {Motome}, \citenamefont {Kwon}, \citenamefont {Adroja},
		\citenamefont {Voneshen}, \citenamefont {Kim}, \citenamefont {Jang} \emph
		{et~al.}}]{do2017}%
	\BibitemOpen
	\bibfield  {author} {\bibinfo {author} {\bibfnamefont {S.-H.}\ \bibnamefont
			{Do}}, \bibinfo {author} {\bibfnamefont {S.-Y.}\ \bibnamefont {Park}},
		\bibinfo {author} {\bibfnamefont {J.}~\bibnamefont {Yoshitake}}, \bibinfo
		{author} {\bibfnamefont {J.}~\bibnamefont {Nasu}}, \bibinfo {author}
		{\bibfnamefont {Y.}~\bibnamefont {Motome}}, \bibinfo {author} {\bibfnamefont
			{Y.~S.}\ \bibnamefont {Kwon}}, \bibinfo {author} {\bibfnamefont
			{D.}~\bibnamefont {Adroja}}, \bibinfo {author} {\bibfnamefont
			{D.}~\bibnamefont {Voneshen}}, \bibinfo {author} {\bibfnamefont
			{K.}~\bibnamefont {Kim}}, \bibinfo {author} {\bibfnamefont {T.-H.}\
			\bibnamefont {Jang}}, \emph {et~al.},\ }\bibfield  {title} {\bibinfo {title}
		{{Majorana fermions in the Kitaev quantum spin system $\alpha$-RuCl$_3$}},\
	}\href@noop {} {\bibfield  {journal} {\bibinfo  {journal} {Nature Physics}\
		}\textbf {\bibinfo {volume} {13}},\ \bibinfo {pages} {1079} (\bibinfo {year}
		{2017})}\BibitemShut {NoStop}%
	\bibitem [{\citenamefont {Laurell}\ and\ \citenamefont
		{Okamoto}(2020)}]{laurell2020}%
	\BibitemOpen
	\bibfield  {author} {\bibinfo {author} {\bibfnamefont {P.}~\bibnamefont
			{Laurell}}\ and\ \bibinfo {author} {\bibfnamefont {S.}~\bibnamefont
			{Okamoto}},\ }\bibfield  {title} {\bibinfo {title} {{Dynamical and thermal
				magnetic properties of the Kitaev spin liquid candidate $\alpha$-RuCl$_3$}},\
	}\href {https://doi.org/10.1038/s41535-019-0203-y} {\bibfield  {journal}
		{\bibinfo  {journal} {npj Quantum Materials}\ }\textbf {\bibinfo {volume}
			{5}},\ \bibinfo {pages} {2} (\bibinfo {year} {2020})}\BibitemShut {NoStop}%
	\bibitem [{\citenamefont {Li}\ \emph {et~al.}(2021)\citenamefont {Li},
		\citenamefont {Zhang}, \citenamefont {Wang}, \citenamefont {Wu},
		\citenamefont {Gao}, \citenamefont {Qu}, \citenamefont {Liu}, \citenamefont
		{Gong},\ and\ \citenamefont {Li}}]{li2021}%
	\BibitemOpen
	\bibfield  {author} {\bibinfo {author} {\bibfnamefont {H.}~\bibnamefont
			{Li}}, \bibinfo {author} {\bibfnamefont {H.-K.}\ \bibnamefont {Zhang}},
		\bibinfo {author} {\bibfnamefont {J.}~\bibnamefont {Wang}}, \bibinfo {author}
		{\bibfnamefont {H.-Q.}\ \bibnamefont {Wu}}, \bibinfo {author} {\bibfnamefont
			{Y.}~\bibnamefont {Gao}}, \bibinfo {author} {\bibfnamefont {D.-W.}\
			\bibnamefont {Qu}}, \bibinfo {author} {\bibfnamefont {Z.-X.}\ \bibnamefont
			{Liu}}, \bibinfo {author} {\bibfnamefont {S.-S.}\ \bibnamefont {Gong}},\ and\
		\bibinfo {author} {\bibfnamefont {W.}~\bibnamefont {Li}},\ }\bibfield
	{title} {\bibinfo {title} {{Identification of magnetic interactions and
				high-field quantum spin liquid in $\alpha$-RuCl$_3$}},\ }\href
	{https://doi.org/10.1038/s41467-021-24257-8} {\bibfield  {journal} {\bibinfo
			{journal} {Nature Communications}\ }\textbf {\bibinfo {volume} {12}},\
		\bibinfo {pages} {4007} (\bibinfo {year} {2021})}\BibitemShut {NoStop}%
	\bibitem [{\citenamefont {Janssen}\ \emph {et~al.}(2020)\citenamefont
		{Janssen}, \citenamefont {Koch},\ and\ \citenamefont {Vojta}}]{janssen2020}%
	\BibitemOpen
	\bibfield  {author} {\bibinfo {author} {\bibfnamefont {L.}~\bibnamefont
			{Janssen}}, \bibinfo {author} {\bibfnamefont {S.}~\bibnamefont {Koch}},\ and\
		\bibinfo {author} {\bibfnamefont {M.}~\bibnamefont {Vojta}},\ }\bibfield
	{title} {\bibinfo {title} {{Magnon dispersion and dynamic spin response in
				three-dimensional spin models for $\alpha$-RuCl$_3$}},\ }\href@noop {}
	{\bibfield  {journal} {\bibinfo  {journal} {Physical Review B}\ }\textbf
		{\bibinfo {volume} {101}},\ \bibinfo {pages} {174444} (\bibinfo {year}
		{2020})}\BibitemShut {NoStop}%
	\bibitem [{\citenamefont {Suzuki}\ \emph {et~al.}(2021)\citenamefont {Suzuki},
		\citenamefont {Liu}, \citenamefont {Bertinshaw}, \citenamefont {Ueda},
		\citenamefont {Kim}, \citenamefont {Laha}, \citenamefont {Weber},
		\citenamefont {Yang}, \citenamefont {Wang}, \citenamefont {Takahashi} \emph
		{et~al.}}]{suzuki2021}%
	\BibitemOpen
	\bibfield  {author} {\bibinfo {author} {\bibfnamefont {H.}~\bibnamefont
			{Suzuki}}, \bibinfo {author} {\bibfnamefont {H.}~\bibnamefont {Liu}},
		\bibinfo {author} {\bibfnamefont {J.}~\bibnamefont {Bertinshaw}}, \bibinfo
		{author} {\bibfnamefont {K.}~\bibnamefont {Ueda}}, \bibinfo {author}
		{\bibfnamefont {H.}~\bibnamefont {Kim}}, \bibinfo {author} {\bibfnamefont
			{S.}~\bibnamefont {Laha}}, \bibinfo {author} {\bibfnamefont {D.}~\bibnamefont
			{Weber}}, \bibinfo {author} {\bibfnamefont {Z.}~\bibnamefont {Yang}},
		\bibinfo {author} {\bibfnamefont {L.}~\bibnamefont {Wang}}, \bibinfo {author}
		{\bibfnamefont {H.}~\bibnamefont {Takahashi}}, \emph {et~al.},\ }\bibfield
	{title} {\bibinfo {title} {{Proximate ferromagnetic state in the Kitaev model
				material $\alpha$-RuCl$_3$}},\ }\href@noop {} {\bibfield  {journal} {\bibinfo
			{journal} {Nature communications}\ }\textbf {\bibinfo {volume} {12}},\
		\bibinfo {pages} {4512} (\bibinfo {year} {2021})}\BibitemShut {NoStop}%
	\bibitem [{\citenamefont {Ritter}(2016)}]{ritter2016}%
	\BibitemOpen
	\bibfield  {author} {\bibinfo {author} {\bibfnamefont {C.}~\bibnamefont
			{Ritter}},\ }\bibfield  {title} {\bibinfo {title} {{Zigzag type magnetic
				structure of the spin $J_{eff}$=$1/2$ compound $\alpha$-RuCl$_3$ as
				determined by neutron powder diffraction}},\ }\href
	{https://doi.org/10.1088/1742-6596/746/1/012060} {\bibfield  {journal}
		{\bibinfo  {journal} {Journal of Physics: Conference Series}\ }\textbf
		{\bibinfo {volume} {746}},\ \bibinfo {pages} {012060} (\bibinfo {year}
		{2016})}\BibitemShut {NoStop}%
	\bibitem [{\citenamefont {Park}\ \emph {et~al.}(2024)\citenamefont {Park},
		\citenamefont {Do}, \citenamefont {Choi}, \citenamefont {Jang}, \citenamefont
		{Jang}, \citenamefont {Scheffer}, \citenamefont {Wu}, \citenamefont
		{Gardner}, \citenamefont {Park}, \citenamefont {Park},\ and\ \citenamefont
		{Ji}}]{park2024}%
	\BibitemOpen
	\bibfield  {author} {\bibinfo {author} {\bibfnamefont {S.-Y.}\ \bibnamefont
			{Park}}, \bibinfo {author} {\bibfnamefont {S.-H.}\ \bibnamefont {Do}},
		\bibinfo {author} {\bibfnamefont {K.-Y.}\ \bibnamefont {Choi}}, \bibinfo
		{author} {\bibfnamefont {D.}~\bibnamefont {Jang}}, \bibinfo {author}
		{\bibfnamefont {T.-H.}\ \bibnamefont {Jang}}, \bibinfo {author}
		{\bibfnamefont {J.}~\bibnamefont {Scheffer}}, \bibinfo {author}
		{\bibfnamefont {C.-M.}\ \bibnamefont {Wu}}, \bibinfo {author} {\bibfnamefont
			{J.~S.}\ \bibnamefont {Gardner}}, \bibinfo {author} {\bibfnamefont
			{J.~M.~S.}\ \bibnamefont {Park}}, \bibinfo {author} {\bibfnamefont {J.-H.}\
			\bibnamefont {Park}},\ and\ \bibinfo {author} {\bibfnamefont
			{S.}~\bibnamefont {Ji}},\ }\bibfield  {title} {\bibinfo {title} {{Emergence
				of the isotropic Kitaev honeycomb lattice $\alpha$-RuCl$_3$ and its magnetic
				properties}},\ }\href {https://doi.org/10.1088/1361-648X/ad294f} {\bibfield
		{journal} {\bibinfo  {journal} {Journal of Physics: Condensed Matter}\
		}\textbf {\bibinfo {volume} {36}},\ \bibinfo {pages} {215803} (\bibinfo
		{year} {2024})}\BibitemShut {NoStop}%
	\bibitem [{\citenamefont {Kim}\ \emph {et~al.}(2024{\natexlab{a}})\citenamefont
		{Kim}, \citenamefont {Horsley}, \citenamefont {Ruff}, \citenamefont
		{Moreno},\ and\ \citenamefont {Kim}}]{kim2024}%
	\BibitemOpen
	\bibfield  {author} {\bibinfo {author} {\bibfnamefont {S.}~\bibnamefont
			{Kim}}, \bibinfo {author} {\bibfnamefont {E.}~\bibnamefont {Horsley}},
		\bibinfo {author} {\bibfnamefont {J.~P.}\ \bibnamefont {Ruff}}, \bibinfo
		{author} {\bibfnamefont {B.~D.}\ \bibnamefont {Moreno}},\ and\ \bibinfo
		{author} {\bibfnamefont {Y.-J.}\ \bibnamefont {Kim}},\ }\bibfield  {title}
	{\bibinfo {title} {{Structural transition and magnetic anisotropy in
				$\alpha$-RuCl$_3$}},\ }\href@noop {} {\bibfield  {journal} {\bibinfo
			{journal} {Physical Review B}\ }\textbf {\bibinfo {volume} {109}},\ \bibinfo
		{pages} {L140101} (\bibinfo {year} {2024}{\natexlab{a}})}\BibitemShut
	{NoStop}%
	\bibitem [{\citenamefont {Sarkis}\ \emph {et~al.}(2024)\citenamefont {Sarkis},
		\citenamefont {Villanova}, \citenamefont {Eichstaedt}, \citenamefont
		{Eguiluz}, \citenamefont {Fernandez-Baca}, \citenamefont {Matsuda},
		\citenamefont {Yan}, \citenamefont {Balz}, \citenamefont {Banerjee},
		\citenamefont {Tennant}, \citenamefont {Berlijn},\ and\ \citenamefont
		{Nagler}}]{sarkis2024}%
	\BibitemOpen
	\bibfield  {author} {\bibinfo {author} {\bibfnamefont {C.~L.}\ \bibnamefont
			{Sarkis}}, \bibinfo {author} {\bibfnamefont {J.~W.}\ \bibnamefont
			{Villanova}}, \bibinfo {author} {\bibfnamefont {C.}~\bibnamefont
			{Eichstaedt}}, \bibinfo {author} {\bibfnamefont {A.~G.}\ \bibnamefont
			{Eguiluz}}, \bibinfo {author} {\bibfnamefont {J.~A.}\ \bibnamefont
			{Fernandez-Baca}}, \bibinfo {author} {\bibfnamefont {M.}~\bibnamefont
			{Matsuda}}, \bibinfo {author} {\bibfnamefont {J.}~\bibnamefont {Yan}},
		\bibinfo {author} {\bibfnamefont {C.}~\bibnamefont {Balz}}, \bibinfo {author}
		{\bibfnamefont {A.}~\bibnamefont {Banerjee}}, \bibinfo {author}
		{\bibfnamefont {D.~A.}\ \bibnamefont {Tennant}}, \bibinfo {author}
		{\bibfnamefont {T.}~\bibnamefont {Berlijn}},\ and\ \bibinfo {author}
		{\bibfnamefont {S.~E.}\ \bibnamefont {Nagler}},\ }\bibfield  {title}
	{\bibinfo {title} {{Experimental evidence for nonspherical magnetic form
				factor in ${\mathrm{Ru}}^{3+}$}},\ }\href
	{https://doi.org/10.1103/PhysRevB.109.104432} {\bibfield  {journal} {\bibinfo
			{journal} {Phys. Rev. B}\ }\textbf {\bibinfo {volume} {109}},\ \bibinfo
		{pages} {104432} (\bibinfo {year} {2024})}\BibitemShut {NoStop}%
	\bibitem [{\citenamefont {Hwan~Chun}\ \emph {et~al.}(2015)\citenamefont
		{Hwan~Chun}, \citenamefont {Kim}, \citenamefont {Kim}, \citenamefont {Zheng},
		\citenamefont {Stoumpos}, \citenamefont {Malliakas}, \citenamefont
		{Mitchell}, \citenamefont {Mehlawat}, \citenamefont {Singh}, \citenamefont
		{Choi}, \citenamefont {Gog}, \citenamefont {Al-Zein}, \citenamefont {Sala},
		\citenamefont {Krisch}, \citenamefont {Chaloupka}, \citenamefont {Jackeli},
		\citenamefont {Khaliullin},\ and\ \citenamefont {Kim}}]{hwanchun2015}%
	\BibitemOpen
	\bibfield  {author} {\bibinfo {author} {\bibfnamefont {S.}~\bibnamefont
			{Hwan~Chun}}, \bibinfo {author} {\bibfnamefont {J.-W.}\ \bibnamefont {Kim}},
		\bibinfo {author} {\bibfnamefont {J.}~\bibnamefont {Kim}}, \bibinfo {author}
		{\bibfnamefont {H.}~\bibnamefont {Zheng}}, \bibinfo {author} {\bibfnamefont
			{C.}~\bibnamefont {Stoumpos}}, \bibinfo {author} {\bibfnamefont {C.~D.}\
			\bibnamefont {Malliakas}}, \bibinfo {author} {\bibfnamefont {J.~F.}\
			\bibnamefont {Mitchell}}, \bibinfo {author} {\bibfnamefont {K.}~\bibnamefont
			{Mehlawat}}, \bibinfo {author} {\bibfnamefont {Y.}~\bibnamefont {Singh}},
		\bibinfo {author} {\bibfnamefont {Y.}~\bibnamefont {Choi}}, \bibinfo {author}
		{\bibfnamefont {T.}~\bibnamefont {Gog}}, \bibinfo {author} {\bibfnamefont
			{A.}~\bibnamefont {Al-Zein}}, \bibinfo {author} {\bibfnamefont {M.~M.}\
			\bibnamefont {Sala}}, \bibinfo {author} {\bibfnamefont {M.}~\bibnamefont
			{Krisch}}, \bibinfo {author} {\bibfnamefont {J.}~\bibnamefont {Chaloupka}},
		\bibinfo {author} {\bibfnamefont {G.}~\bibnamefont {Jackeli}}, \bibinfo
		{author} {\bibfnamefont {G.}~\bibnamefont {Khaliullin}},\ and\ \bibinfo
		{author} {\bibfnamefont {B.~J.}\ \bibnamefont {Kim}},\ }\bibfield  {title}
	{\bibinfo {title} {{Direct evidence for dominant bond-directional
				interactions in a honeycomb lattice iridate Na$_2$IrO$_3$}},\ }\href
	{https://doi.org/10.1038/nphys3322} {\bibfield  {journal} {\bibinfo
			{journal} {Nature Physics}\ }\textbf {\bibinfo {volume} {11}},\ \bibinfo
		{pages} {462} (\bibinfo {year} {2015})}\BibitemShut {NoStop}%
	\bibitem [{\citenamefont {Magnaterra}\ \emph {et~al.}(2023)\citenamefont
		{Magnaterra}, \citenamefont {Hopfer}, \citenamefont {Sahle}, \citenamefont
		{Sala}, \citenamefont {Monaco}, \citenamefont {Attig}, \citenamefont
		{Hickey}, \citenamefont {Pietsch}, \citenamefont {Breitner}, \citenamefont
		{Gegenwart} \emph {et~al.}}]{magnaterra2023}%
	\BibitemOpen
	\bibfield  {author} {\bibinfo {author} {\bibfnamefont {M.}~\bibnamefont
			{Magnaterra}}, \bibinfo {author} {\bibfnamefont {K.}~\bibnamefont {Hopfer}},
		\bibinfo {author} {\bibfnamefont {C.~J.}\ \bibnamefont {Sahle}}, \bibinfo
		{author} {\bibfnamefont {M.~M.}\ \bibnamefont {Sala}}, \bibinfo {author}
		{\bibfnamefont {G.}~\bibnamefont {Monaco}}, \bibinfo {author} {\bibfnamefont
			{J.}~\bibnamefont {Attig}}, \bibinfo {author} {\bibfnamefont
			{C.}~\bibnamefont {Hickey}}, \bibinfo {author} {\bibfnamefont {I.-M.}\
			\bibnamefont {Pietsch}}, \bibinfo {author} {\bibfnamefont {F.}~\bibnamefont
			{Breitner}}, \bibinfo {author} {\bibfnamefont {P.}~\bibnamefont {Gegenwart}},
		\emph {et~al.},\ }\bibfield  {title} {\bibinfo {title} {{RIXS observation of
				bond-directional nearest-neighbor excitations in the Kitaev material
				Na$_2$IrO$_3$}},\ }\href@noop {} {\bibfield  {journal} {\bibinfo  {journal}
			{arXiv preprint arXiv:2301.08340}\ } (\bibinfo {year} {2023})}\BibitemShut
	{NoStop}%
	\bibitem [{\citenamefont {Chaloupka}\ and\ \citenamefont
		{Khaliullin}(2015)}]{chaloupka2015}%
	\BibitemOpen
	\bibfield  {author} {\bibinfo {author} {\bibfnamefont {J.}~\bibnamefont
			{Chaloupka}}\ and\ \bibinfo {author} {\bibfnamefont {G.}~\bibnamefont
			{Khaliullin}},\ }\bibfield  {title} {\bibinfo {title} {{Hidden symmetries of
				the extended Kitaev-Heisenberg model: Implications for the honeycomb-lattice
				iridates ${A}_{2}{\mathrm{IrO}}_{3}$}},\ }\href
	{https://doi.org/10.1103/PhysRevB.92.024413} {\bibfield  {journal} {\bibinfo
			{journal} {Phys. Rev. B}\ }\textbf {\bibinfo {volume} {92}},\ \bibinfo
		{pages} {024413} (\bibinfo {year} {2015})}\BibitemShut {NoStop}%
	\bibitem [{\citenamefont {Maksimov}\ and\ \citenamefont
		{Chernyshev}(2020)}]{maksimov2020}%
	\BibitemOpen
	\bibfield  {author} {\bibinfo {author} {\bibfnamefont {P.~A.}\ \bibnamefont
			{Maksimov}}\ and\ \bibinfo {author} {\bibfnamefont {A.~L.}\ \bibnamefont
			{Chernyshev}},\ }\bibfield  {title} {\bibinfo {title} {{Rethinking
				$\ensuremath{\alpha}\text{\ensuremath{-}}{\mathrm{RuCl}}_{3}$}},\ }\href
	{https://doi.org/10.1103/PhysRevResearch.2.033011} {\bibfield  {journal}
		{\bibinfo  {journal} {Phys. Rev. Res.}\ }\textbf {\bibinfo {volume} {2}},\
		\bibinfo {pages} {033011} (\bibinfo {year} {2020})}\BibitemShut {NoStop}%
	\bibitem [{\citenamefont {Chatterji}(2005)}]{chatterji2005}%
	\BibitemOpen
	\bibfield  {author} {\bibinfo {author} {\bibfnamefont {T.}~\bibnamefont
			{Chatterji}},\ }\href@noop {} {\emph {\bibinfo {title} {Neutron scattering
				from magnetic materials}}}\ (\bibinfo  {publisher} {Elsevier},\ \bibinfo
	{year} {2005})\BibitemShut {NoStop}%
	\bibitem [{\citenamefont {Qureshi}\ \emph {et~al.}(2012)\citenamefont
		{Qureshi}, \citenamefont {Steffens}, \citenamefont {Wurmehl}, \citenamefont
		{Aswartham}, \citenamefont {B\"uchner},\ and\ \citenamefont
		{Braden}}]{qureshi2012}%
	\BibitemOpen
	\bibfield  {author} {\bibinfo {author} {\bibfnamefont {N.}~\bibnamefont
			{Qureshi}}, \bibinfo {author} {\bibfnamefont {P.}~\bibnamefont {Steffens}},
		\bibinfo {author} {\bibfnamefont {S.}~\bibnamefont {Wurmehl}}, \bibinfo
		{author} {\bibfnamefont {S.}~\bibnamefont {Aswartham}}, \bibinfo {author}
		{\bibfnamefont {B.}~\bibnamefont {B\"uchner}},\ and\ \bibinfo {author}
		{\bibfnamefont {M.}~\bibnamefont {Braden}},\ }\bibfield  {title} {\bibinfo
		{title} {{Local magnetic anisotropy in BaFe${}_{2}$As${}_{2}$: A polarized
				inelastic neutron scattering study}},\ }\href
	{https://doi.org/10.1103/PhysRevB.86.060410} {\bibfield  {journal} {\bibinfo
			{journal} {Phys. Rev. B}\ }\textbf {\bibinfo {volume} {86}},\ \bibinfo
		{pages} {060410} (\bibinfo {year} {2012})}\BibitemShut {NoStop}%
	\bibitem [{\citenamefont {Kunkem\"oller}\ \emph {et~al.}(2015)\citenamefont
		{Kunkem\"oller}, \citenamefont {Khomskii}, \citenamefont {Steffens},
		\citenamefont {Piovano}, \citenamefont {Nugroho},\ and\ \citenamefont
		{Braden}}]{kunkemoeller2015}%
	\BibitemOpen
	\bibfield  {author} {\bibinfo {author} {\bibfnamefont {S.}~\bibnamefont
			{Kunkem\"oller}}, \bibinfo {author} {\bibfnamefont {D.}~\bibnamefont
			{Khomskii}}, \bibinfo {author} {\bibfnamefont {P.}~\bibnamefont {Steffens}},
		\bibinfo {author} {\bibfnamefont {A.}~\bibnamefont {Piovano}}, \bibinfo
		{author} {\bibfnamefont {A.~A.}\ \bibnamefont {Nugroho}},\ and\ \bibinfo
		{author} {\bibfnamefont {M.}~\bibnamefont {Braden}},\ }\bibfield  {title}
	{\bibinfo {title} {{Highly Anisotropic Magnon Dispersion in
				${\mathrm{Ca}}_{2}{\mathrm{RuO}}_{4}$: Evidence for Strong Spin Orbit
				Coupling}},\ }\href {https://doi.org/10.1103/PhysRevLett.115.247201}
	{\bibfield  {journal} {\bibinfo  {journal} {Phys. Rev. Lett.}\ }\textbf
		{\bibinfo {volume} {115}},\ \bibinfo {pages} {247201} (\bibinfo {year}
		{2015})}\BibitemShut {NoStop}%
	\bibitem [{\citenamefont {Mi}\ \emph {et~al.}(2021)\citenamefont {Mi},
		\citenamefont {Wang}, \citenamefont {Gui}, \citenamefont {Pi}, \citenamefont
		{Zheng}, \citenamefont {Yang}, \citenamefont {Gan}, \citenamefont {Wang},
		\citenamefont {Li}, \citenamefont {Wang}, \citenamefont {Zhang},
		\citenamefont {Su}, \citenamefont {Chai},\ and\ \citenamefont {He}}]{mi2021}%
	\BibitemOpen
	\bibfield  {author} {\bibinfo {author} {\bibfnamefont {X.}~\bibnamefont
			{Mi}}, \bibinfo {author} {\bibfnamefont {X.}~\bibnamefont {Wang}}, \bibinfo
		{author} {\bibfnamefont {H.}~\bibnamefont {Gui}}, \bibinfo {author}
		{\bibfnamefont {M.}~\bibnamefont {Pi}}, \bibinfo {author} {\bibfnamefont
			{T.}~\bibnamefont {Zheng}}, \bibinfo {author} {\bibfnamefont
			{K.}~\bibnamefont {Yang}}, \bibinfo {author} {\bibfnamefont {Y.}~\bibnamefont
			{Gan}}, \bibinfo {author} {\bibfnamefont {P.}~\bibnamefont {Wang}}, \bibinfo
		{author} {\bibfnamefont {A.}~\bibnamefont {Li}}, \bibinfo {author}
		{\bibfnamefont {A.}~\bibnamefont {Wang}}, \bibinfo {author} {\bibfnamefont
			{L.}~\bibnamefont {Zhang}}, \bibinfo {author} {\bibfnamefont
			{Y.}~\bibnamefont {Su}}, \bibinfo {author} {\bibfnamefont {Y.}~\bibnamefont
			{Chai}},\ and\ \bibinfo {author} {\bibfnamefont {M.}~\bibnamefont {He}},\
	}\bibfield  {title} {\bibinfo {title} {{Stacking faults in
				$\ensuremath{\alpha}\text{\ensuremath{-}}{\mathrm{RuCl}}_{3}$ revealed by
				local electric polarization}},\ }\href
	{https://doi.org/10.1103/PhysRevB.103.174413} {\bibfield  {journal} {\bibinfo
			{journal} {Phys. Rev. B}\ }\textbf {\bibinfo {volume} {103}},\ \bibinfo
		{pages} {174413} (\bibinfo {year} {2021})}\BibitemShut {NoStop}%
	\bibitem [{\citenamefont {Bertin}\ \emph {et~al.}(2023)\citenamefont {Bertin},
		\citenamefont {Braden}, \citenamefont {Steffens},\ and\ \citenamefont
		{Su}}]{a-rcl_thales}%
	\BibitemOpen
	\bibfield  {author} {\bibinfo {author} {\bibfnamefont {A.}~\bibnamefont
			{Bertin}}, \bibinfo {author} {\bibfnamefont {M.}~\bibnamefont {Braden}},
		\bibinfo {author} {\bibfnamefont {P.}~\bibnamefont {Steffens}},\ and\
		\bibinfo {author} {\bibfnamefont {Y.}~\bibnamefont {Su}},\ }\href@noop {} {}
	(\bibinfo {year} {2023}),\ \bibinfo {note} {{ "Study of bond-directional
			excitations in the proximate Kitaev QSL $\alpha$-RuCl$_3$". Institut
			Laue-Langevin (ILL) doi:10.5291/ILL-DATA.4-01-1789.}}\BibitemShut {Stop}%
	\bibitem [{\citenamefont {Sears}\ \emph {et~al.}(2020)\citenamefont {Sears},
		\citenamefont {Chern}, \citenamefont {Kim}, \citenamefont {Bereciartua},
		\citenamefont {Francoual}, \citenamefont {Kim},\ and\ \citenamefont
		{Kim}}]{sears2020}%
	\BibitemOpen
	\bibfield  {author} {\bibinfo {author} {\bibfnamefont {J.~A.}\ \bibnamefont
			{Sears}}, \bibinfo {author} {\bibfnamefont {L.~E.}\ \bibnamefont {Chern}},
		\bibinfo {author} {\bibfnamefont {S.}~\bibnamefont {Kim}}, \bibinfo {author}
		{\bibfnamefont {P.~J.}\ \bibnamefont {Bereciartua}}, \bibinfo {author}
		{\bibfnamefont {S.}~\bibnamefont {Francoual}}, \bibinfo {author}
		{\bibfnamefont {Y.~B.}\ \bibnamefont {Kim}},\ and\ \bibinfo {author}
		{\bibfnamefont {Y.-J.}\ \bibnamefont {Kim}},\ }\bibfield  {title} {\bibinfo
		{title} {{Ferromagnetic Kitaev interaction and the origin of large magnetic
				anisotropy in $\alpha$-{R}u{C}l$_{3}$}},\ }\href
	{https://doi.org/10.1038/s41567-020-0874-0} {\bibfield  {journal} {\bibinfo
			{journal} {Nature physics}\ }\textbf {\bibinfo {volume} {16}},\ \bibinfo
		{pages} {837 – 840} (\bibinfo {year} {2020})}\BibitemShut {NoStop}%
	\bibitem [{\citenamefont {Kim}\ \emph {et~al.}(2024{\natexlab{b}})\citenamefont
		{Kim}, \citenamefont {Kim}, \citenamefont {Horsley}, \citenamefont {Nelson},\
		and\ \citenamefont {Ruff}}]{kim2024b}%
	\BibitemOpen
	\bibfield  {author} {\bibinfo {author} {\bibfnamefont {Y.-J.}\ \bibnamefont
			{Kim}}, \bibinfo {author} {\bibfnamefont {S.}~\bibnamefont {Kim}}, \bibinfo
		{author} {\bibfnamefont {E.}~\bibnamefont {Horsley}}, \bibinfo {author}
		{\bibfnamefont {C.}~\bibnamefont {Nelson}},\ and\ \bibinfo {author}
		{\bibfnamefont {J.}~\bibnamefont {Ruff}},\ }\bibfield  {title} {\bibinfo
		{title} {{Re-investigation of Moment Direction in a Kitaev Material
				$\alpha$-RuCl$_3$}},\ }\bibfield  {journal} {\bibinfo  {journal} {Research
			Square}\ }\href {https://doi.org/https://doi.org/10.21203/rs.3.rs-3894945/v1}
	{https://doi.org/10.21203/rs.3.rs-3894945/v1} (\bibinfo {year}
	{2024}{\natexlab{b}})\BibitemShut {NoStop}%
	\bibitem [{\citenamefont {{Le Guillou, J.C.}}\ and\ \citenamefont
		{{Zinn-Justin, J.}}(1987)}]{leguillou1987}%
	\BibitemOpen
	\bibfield  {author} {\bibinfo {author} {\bibnamefont {{Le Guillou, J.C.}}}\
		and\ \bibinfo {author} {\bibnamefont {{Zinn-Justin, J.}}},\ }\bibfield
	{title} {\bibinfo {title} {Accurate critical exponents for ising like systems
			in non-integer dimensions},\ }\href
	{https://doi.org/10.1051/jphys:0198700480101900} {\bibfield  {journal}
		{\bibinfo  {journal} {J. Phys. France}\ }\textbf {\bibinfo {volume} {48}},\
		\bibinfo {pages} {19} (\bibinfo {year} {1987})}\BibitemShut {NoStop}%
	\bibitem [{\citenamefont {Novotny}(1992)}]{novotny1992}%
	\BibitemOpen
	\bibfield  {author} {\bibinfo {author} {\bibfnamefont {M.~A.}\ \bibnamefont
			{Novotny}},\ }\bibfield  {title} {\bibinfo {title} {Critical exponents for
			the ising model between one and two dimensions},\ }\href
	{https://doi.org/10.1103/PhysRevB.46.2939} {\bibfield  {journal} {\bibinfo
			{journal} {Phys. Rev. B}\ }\textbf {\bibinfo {volume} {46}},\ \bibinfo
		{pages} {2939} (\bibinfo {year} {1992})}\BibitemShut {NoStop}%
	\bibitem [{\citenamefont {Toth}\ and\ \citenamefont {Lake}(2015)}]{toth2015}%
	\BibitemOpen
	\bibfield  {author} {\bibinfo {author} {\bibfnamefont {S.}~\bibnamefont
			{Toth}}\ and\ \bibinfo {author} {\bibfnamefont {B.}~\bibnamefont {Lake}},\
	}\bibfield  {title} {\bibinfo {title} {Linear spin wave theory for single-q
			incommensurate magnetic structures},\ }\href
	{https://doi.org/10.1088/0953-8984/27/16/166002} {\bibfield  {journal}
		{\bibinfo  {journal} {Journal of Physics: Condensed Matter}\ }\textbf
		{\bibinfo {volume} {27}},\ \bibinfo {pages} {166002} (\bibinfo {year}
		{2015})}\BibitemShut {NoStop}%
	\bibitem [{sup()}]{suppl-mat}%
	\BibitemOpen
	\href@noop {} {}\bibinfo {note} {Supplemental information about the linear
		spin-wave theory calculations is given in *****}\BibitemShut {NoStop}%
	\bibitem [{\citenamefont {Jain}\ \emph {et~al.}(2017)\citenamefont {Jain},
		\citenamefont {Krautloher}, \citenamefont {Porras}, \citenamefont {Ryu},
		\citenamefont {Chen}, \citenamefont {Abernathy}, \citenamefont {Park},
		\citenamefont {Ivanov}, \citenamefont {Chaloupka}, \citenamefont {Khaliullin}
		\emph {et~al.}}]{jain2017}%
	\BibitemOpen
	\bibfield  {author} {\bibinfo {author} {\bibfnamefont {A.}~\bibnamefont
			{Jain}}, \bibinfo {author} {\bibfnamefont {M.}~\bibnamefont {Krautloher}},
		\bibinfo {author} {\bibfnamefont {J.}~\bibnamefont {Porras}}, \bibinfo
		{author} {\bibfnamefont {G.}~\bibnamefont {Ryu}}, \bibinfo {author}
		{\bibfnamefont {D.}~\bibnamefont {Chen}}, \bibinfo {author} {\bibfnamefont
			{D.}~\bibnamefont {Abernathy}}, \bibinfo {author} {\bibfnamefont
			{J.}~\bibnamefont {Park}}, \bibinfo {author} {\bibfnamefont {A.}~\bibnamefont
			{Ivanov}}, \bibinfo {author} {\bibfnamefont {J.}~\bibnamefont {Chaloupka}},
		\bibinfo {author} {\bibfnamefont {G.}~\bibnamefont {Khaliullin}}, \emph
		{et~al.},\ }\bibfield  {title} {\bibinfo {title} {Higgs mode and its decay in
			a two-dimensional antiferromagnet},\ }\href@noop {} {\bibfield  {journal}
		{\bibinfo  {journal} {Nature Physics}\ }\textbf {\bibinfo {volume} {13}},\
		\bibinfo {pages} {633} (\bibinfo {year} {2017})}\BibitemShut {NoStop}%
	\bibitem [{\citenamefont {Kunkem{\"o}ller}\ \emph {et~al.}(2017)\citenamefont
		{Kunkem{\"o}ller}, \citenamefont {Komleva}, \citenamefont {Streltsov},
		\citenamefont {Hoffmann}, \citenamefont {Khomskii}, \citenamefont {Steffens},
		\citenamefont {Sidis}, \citenamefont {Schmalzl},\ and\ \citenamefont
		{Braden}}]{kunkemoeller2017}%
	\BibitemOpen
	\bibfield  {author} {\bibinfo {author} {\bibfnamefont {S.}~\bibnamefont
			{Kunkem{\"o}ller}}, \bibinfo {author} {\bibfnamefont {E.}~\bibnamefont
			{Komleva}}, \bibinfo {author} {\bibfnamefont {S.}~\bibnamefont {Streltsov}},
		\bibinfo {author} {\bibfnamefont {S.}~\bibnamefont {Hoffmann}}, \bibinfo
		{author} {\bibfnamefont {D.}~\bibnamefont {Khomskii}}, \bibinfo {author}
		{\bibfnamefont {P.}~\bibnamefont {Steffens}}, \bibinfo {author}
		{\bibfnamefont {Y.}~\bibnamefont {Sidis}}, \bibinfo {author} {\bibfnamefont
			{K.}~\bibnamefont {Schmalzl}},\ and\ \bibinfo {author} {\bibfnamefont
			{M.}~\bibnamefont {Braden}},\ }\bibfield  {title} {\bibinfo {title} {{Magnon
				dispersion in Ca$_2$Ru$_{1-x}$Ti$_x$O$_4$: Impact of spin-orbit coupling and
				oxygen moments}},\ }\href@noop {} {\bibfield  {journal} {\bibinfo  {journal}
			{Physical Review B}\ }\textbf {\bibinfo {volume} {95}},\ \bibinfo {pages}
		{214408} (\bibinfo {year} {2017})}\BibitemShut {NoStop}%
	\bibitem [{\citenamefont {Braden}\ \emph {et~al.}(2024)\citenamefont {Braden},
		\citenamefont {Bertin}, \citenamefont {Hansen}, \citenamefont {Steffens},\
		and\ \citenamefont {Su}}]{a-rcl_in20}%
	\BibitemOpen
	\bibfield  {author} {\bibinfo {author} {\bibfnamefont {M.}~\bibnamefont
			{Braden}}, \bibinfo {author} {\bibfnamefont {A.}~\bibnamefont {Bertin}},
		\bibinfo {author} {\bibfnamefont {U.~B.}\ \bibnamefont {Hansen}}, \bibinfo
		{author} {\bibfnamefont {P.}~\bibnamefont {Steffens}},\ and\ \bibinfo
		{author} {\bibfnamefont {Y.}~\bibnamefont {Su}},\ }\href@noop {} {} (\bibinfo
	{year} {2024}),\ \bibinfo {note} {{IN20 "Study of bond-directional anisotropy
			in the proximate Kitaev QSL $\alpha$-RuCl$_3$". Institut Laue-Langevin (ILL)
			doi:10.5291/ILL-DATA.4-05-907.}}\BibitemShut {Stop}%
\end{thebibliography}
%

\clearpage


\section*{Supplementary material}

# Direct evidence for anisotropic magnetic interaction in $\alpha$-RuCl$_3$ from polarized inelastic neutron scattering

Markus Braden,[1, *] Xiao Wang,[2, 3, 4] Alexandre Bertin,[1] Paul Steffens,[5] and Yixi Su[2, †]

[1] *II. Physikalisches Institut, Universität zu Köln, Zülpicher Str. 77, D-50937 Köln, Germany*
[2] *Jülich Centre for Neutron Science JCNS at MLZ,
Forschungszentrum Jülich, Lichtenbergstr. 1, D-85747 Garching, Germany*
[3] *Institute of High Energy Physics, Chinese Academy of Sciences (CAS), Beijing 100049, China*
[4] *Spallation Neutron Source Science Center, Dongguan 523803, China*
[5] *Institut Laue-Langevin, 71 avenue des Martyrs, CS 20156, 38042 Grenoble Cedex 9, France*

(Dated: September 13, 2024)

## MAGNON DISPERSION CALCULATED WITH LINEAR SPIN-WAVE THEORY

In spite of tremendous efforts there is no widely accepted set of microscopic interaction parameters to describe magnetic excitations in $\alpha$-RuCl$_3$, and also from the experiment side the observation of well-defined magnons seems too limited to fully establish the magnon dispersion in the ordered state. Sharp magnon modes were unambiguously observed only for the low-lying branch connecting the antiferromagnetic zone center $(0,1,Q_L)$ to the ferromagnetic propagation vector $(0,0,Q_L)$[1] and most experiments focused on the continuum or multi-magnon scattering at higher temperature or at magnetic fields that, however, is little structured in energy and $q$ space. The comparison of the magnon dispersion extracted by linear spin-wave theory and the scattering function determined by exact diagonalization [2] yields similar results validating the spin-wave approach to analyze the microscopic parameters for $\alpha$-RuCl$_3$.

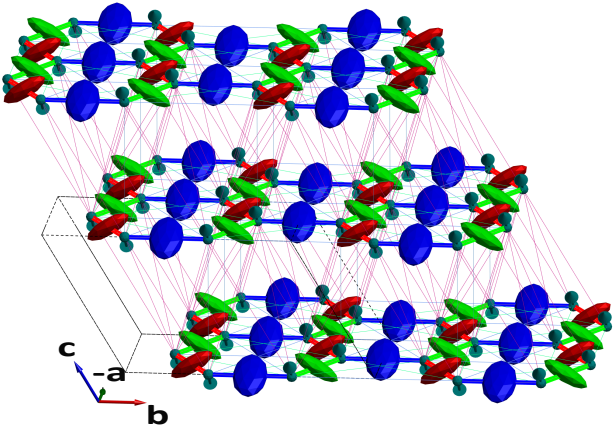

FIG. 1. Picture of the honeycomb arrangement of $\alpha$-RuCl$_3$ including the various magnetic interaction parameters explained in the main text. The anisotropic bond-directional nearest neighbor interaction is illustrated by ellipsoids and various Heisenberg interactions are drawn as differently colored lines. Two types of inter-layer interactions, $J_p$ and $J_p'$ are included.

We have implemented the extended Kitaev Heisenberg model using the the *SpinW* program [3]. We describe the next-nearest neighbor interaction with the the Kitaev interaction $J_K$, a Heisenberg term $J_1$ and two parameters for the symmetric anisotropic interaction, $\Gamma$ and $\Gamma'$ [2, 4, 5] as explained in the main text. Further parameters are the isotropic interaction terms between second- and third-nearest neighbors, $J_2$ and $J_3$. With the model parameters proposed by Winter et al. [2], ferromagnetic $J_K$=-5 meV, ferromagnetic $J_1$=-0.5 meV $\Gamma$=2.5 meV and antiferromagnetic $J_3$=0.5, we perfectly reproduce the calculations in reference [2]. Figure 1 illustrates the different magnetic interaction parameters in the crystal structure of $\alpha$-RuCl$_3$. For numerical reasons we use a non-conventional monoclinic setting in which the $c$ parameter connects neighboring planes. In order to compare with the experimental vertical dispersion a correction factor 3 applies. In the spin-wave model we also include magnetic interaction perpendicular to RuCl$_3$ layers $J_p$ and $J_p'$. In many analyses the neutron data simulation requires to add the contributions of the different magnetic domain orientations (rotated by $\pm$ 120 degrees against each other), which, however, hides the most important aspect to understand our polarized inelastic neutron scattering experiments. Therefore, Fig. 2 and 3 present the spin-wave dispersion and the scattering function without this domain folding in the orthorhombic setting. Ignoring the out-of-plane correlations the magnetic ordering breaks the $C$ centering of a single $a,b$ plane, therefore $(1,0,0)$ and $(0,1,0)$ become magnetic zone centers. The results presented in Fig. 2 also include the finite interlayer interaction parameters, which will be discussed below but are irrelevant for the main conclusions. Without the folding of other domains the simple model does not yield low-energy response at the antiferromagnetic zone center $(0,1,Q_L)$. The usual low-energy transversal magnons that without any magnetic anisotropy linearly disperse from $E$=0 at the magnetic Bragg position and that obtain some finite but small energy for moderate anisotropy are fully suppressed. These modes are shifted to high energies in perfect agreement with the experiment. Note that a simple planar anisotropy would still

yield one low-energy non-degenerate transversal magnon at the Γ point.

The microscopic model for α-RuCl$_3$ with ferromagnetic Kitaev and ferromagnetic Heisenberg $J_1$ interaction is peculiar concerning its closeness to a ferromagnetic instability, see also the discussion in reference [5]. Experimentally, this manifests itself in the sharp and strong low-energy signal observed at (0,0,$Q_L$) or equivalent. The zigzag and ferromagnetic ordering schemes share the ferromagnetic zigzag chain (running vertically in the illustration of Fig. 2a) in the main text) and thus only differ in the antiferromagnetic or ferromagnetic alignment in the bond vertical to the zigzag chains (horizontal in the same Fig. 2a)). The small energy difference between the two ordering schemes results from the interplay of ferromagnetic Heisenberg $J_1$ and antiferromagnetic $J_3$, because by orienting the moments correspondingly the zigzag structure avoids the increase of energy in this bond otherwise induced by the strong ferromagnetic Kitaev term. The low-energy neutron scattering response appearing at antiferromagnetic (0,1,$Q_L$) stems from the rotated domains. It corresponds to the low-energy signal at (0.5,0.5,$Q_L$) from the two magnetic domains with the zigzag chains rotated by ±120 degrees, as it is illustrated in Fig. 2 a-c) in the main text. The anisotropic character of this signal perfectly confirms the bond-directional interaction.

In Fig. 2 a)-d) we compare three recent models from the literature: $MW$ is model 2 in Winter et al. [2], $MM$ is model 1 in Maksimov and Chernyshev [6], $ML$ is taken from Liu et al. [5] and $MO$ is an optimized version of the latter. The parameter values (given in meV) are shown in the title of the panels and document the large spread of parameter sets used in these models. Overall, the dispersion plots and the intensity distribution look nevertheless very similar. One recognizes the most visible steep dispersion along the path from (0,0,0) to (1,0,0), which senses the strong ferromagnetic interaction within the ferromagnetic zigzag segments. Differences concern the width of the dispersion and the splitting between some branches. The models in Fig. 2 a)-c) already contain some extra parameters to improve the description of the neutron data, as will be discussed below.

Our polarized data confirm the finite $Q_L$ dispersion of the low-energy sharp signal appearing at the ferromagnetic zone center. However, we can also follow the $Q_L$ dispersion at (0,1,$Q_L$) finding a much lower dispersion than that calculated in reference [7]. The finite dispersion perpendicular to the RuCl$_3$ layers unambiguously reveals significant interlayer interaction. The two shortest Ru-Ru interlayer distances are included in Fig. 1. There is one distance perfectly perpendicular to the layers that was used in reference [7] to model the vertical dispersion. However a rather large value is needed to model dispersion at (0,0,$Q_L$) that more importantly also generates a strong dispersion of the antiferromagnetic zigzag signal at (0,1,$Q_L$) in disagreement with our experiment. Note that there were no data for this dispersion before our experiment. By using a smaller value for the nextnearest interlayer bond, $J'_p$=0.04 meV, we can satisfactorily describe both vertical dispersion relations, in particular the stronger impact for the quasiferromagnetic signal. This equally holds for the three literature models studied [2, 5, 6], see Fig. 2 a-c). The vertical dispersion of the ferromagnetic signal can be best seen in the path from (0,0,1) to (0,0,0) and the dispersion of the low-energy antiferromagnetic zigzag signal is visible in the path from (0.5,0.5,0) to (0.5,0.5,1). While $J_p$ couples each Ru moment only to one other moment, there are nine such neighbors for $J'_p$ as it can be seen in Fig. 3. The $J'_p$ parameter represents a much smaller correction of the essentially layered magnetic model for α-RuCl$_3$.

The second improvement of the models concerns the low-energy antiferromagnetic mode observed at (0,1,$Q_L$) that is attributed to (0.5,0.5,$Q_L$), i.e. the non-condensed zigzag instabilities. The models yield nearly vanishing energies while the experiment finds ∼2 meV with only weak vertical dispersion. The mode can be stabilized either by modifying the Γ value or by inducing an impact of the orthorhombic magnetic structure on the nearest neighbor interaction parameters. Such an effect is primordial in the antiferromagnetic state of the FeAs parent compounds associated with the nematic character of the magnetic correlations. We introduce a modulation of the ferromagnetic Kitaev terms by $(1-2\delta)K$ for bonds perpendicular to the zigzag chain and by $(1+\delta)K$ parallel. This modulation stabilizes the static magnetic arrangement while increasing the energy of non-condensed modes. An only small modulation of $\delta$=0.04 yields good agreement for all three models.

The lowest transversal magnon at the antiferromagnetic scattering vector is calculated at energies 3.5, 3, and 3.3 meV in the models $MW$, $MM$, and $ML$, respectively, which disagrees with the absence of such a signal in our polarized experiments, see main text. The data in Fig. 3 a) in the main article suggest a signal around 4.5 meV, which is supported by unpolarized data [8] and by a polarized experiment on a thermal spectrometer [9]. The calculated energy of this mode and the total width of the dispersion can be increased in model $ML$ by multiplying $J$ values by 1.3 and Γ and Γ' by 1.5. Since this also enhances the energy of the ferromagnetic mode the latter was softened by adding a ferromagnetic $J_2$=-0.1 meV. The calculations with this model $MO$ are shown Fig. 2d) for the total neutron scattering signal and in Fig. 3 a)-c) for the anisotropic components. The scattering distribution of these components is qualitatively the same for these four models.

Model $MO$ captures much of the experimental results as described above but the comparison suffers from the lack of well defined magnon modes at higher energies, which would be needed to fix the large number of pa-

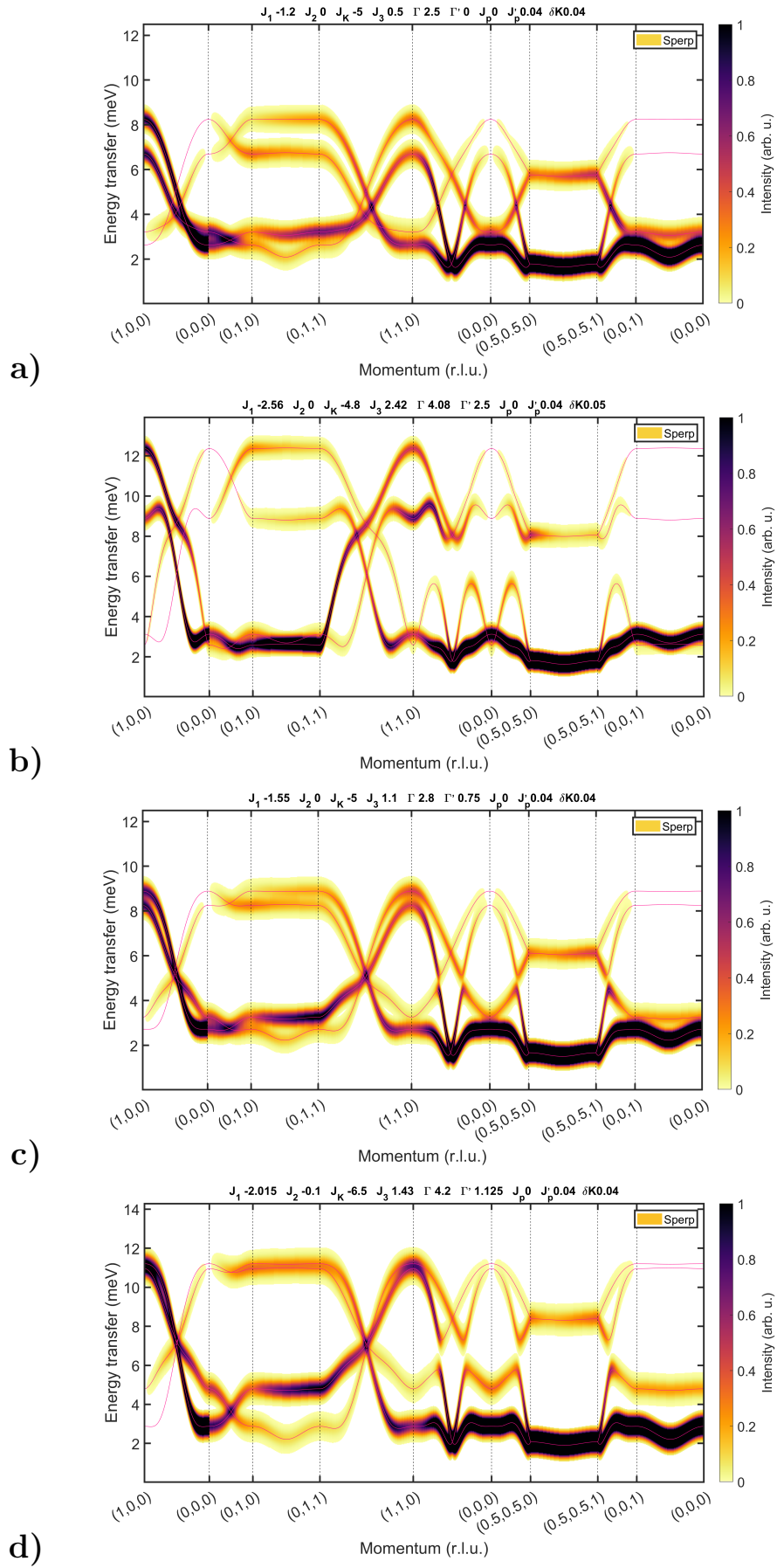

FIG. 2. Magnon dispersion in the zigzag ordered state calculated with the *SpinW* program [3]. The dispersion is drawn by thin lines and the color coding indicates the strength of the total neutron scattering signal, i.e. modes polarized perpendicular to the scattering vector. In the panels a)-d) results for models $MW$ [2], $MM$ [6], $ML$ [5] and an adapted model $MO$ from the latter are presented, respectively, see text.

rameters. For higher energies only broad signals are reported which can be interpreted either as a fingerprint of the neighboring quantum spin-liquid state or as a result from magnon decay due to the fully anisotropic interaction [2, 6].

---

* braden@ph2.uni-koeln.de
† y.su@fz-juelich.de
[1] Note that $(h,k,Q_L)$ with $h+k$ even are equivalent besides the 3D stacking.
[2] S. M. Winter, K. Riedl, P. A. Maksimov, A. L. Chernyshev, A. Honecker, and R. Valentí, Breakdown of magnons in a strongly spin-orbital coupled magnet, Nature Communications **8**, 1152 (2017).
[3] S. Toth and B. Lake, Linear spin wave theory for single-q incommensurate magnetic structures, Journal of Physics: Condensed Matter **27**, 166002 (2015).
[4] S. M. Winter, A. A. Tsirlin, M. Daghofer, J. van den Brink, Y. Singh, P. Gegenwart, and R. Valentí, Models and materials for generalized Kitaev magnetism, J. Phys.: Cond. Matt. **29**, 493002 (2017).
[5] H. Liu, J. Chaloupka, and G. Khaliullin, Exchange interactions in $d^5$ Kitaev materials: From $Na_2IrO_3$ to $\alpha$-$RuCl_3$, Phys. Rev. B **105**, 214411 (2022).
[6] P. A. Maksimov and A. L. Chernyshev, Rethinking $\alpha-RuCl_3$, Phys. Rev. Res. **2**, 033011 (2020).
[7] L. Janssen, S. Koch, and M. Vojta, Magnon dispersion and dynamic spin response in three-dimensional spin models for $\alpha$-$RuCl_3$, Physical Review B **101**, 174444 (2020).
[8] A. Banerjee, P. Lampen-Kelley, J. Knolle, C. Balz, A. A. Aczel, B. Winn, Y. Liu, D. Pajerowski, J. Yan, C. A. Bridges, A. T. Savici, B. C. Chakoumakos, M. D. Lumsden, D. A. Tennant, R. Moessner, D. G. Mandrus, and S. E. Nagler, Excitations in the field-induced quantum spin liquid state of $\alpha$-$RuCl_3$, npj Quantum Materials **3**, 8 (2018).
[9] M. Braden, A. Bertin, U. B. Hansen, P. Steffens, and Y. Su, (2024), IN20 "Study of bond-directional anisotropy in the proximate Kitaev QSL $\alpha$-$RuCl_3$". Institut Laue-Langevin (ILL) doi:10.5291/ILL-DATA.4-05-907.



\end{document}